\documentclass[11pt]{article}

\usepackage[utf8]{inputenc}
\usepackage[T1]{fontenc}
\usepackage{lmodern}

\usepackage{amsmath,amssymb}
\usepackage{mathtools}   
\usepackage{bm}          
\usepackage{amsthm}      

\usepackage{siunitx}     

\usepackage{booktabs}    
\usepackage{array}       
\usepackage{multirow}    
\usepackage{tabularx}    
\usepackage{longtable}   

\usepackage{graphicx}
\usepackage{epstopdf}    
\usepackage{subcaption}
\usepackage[labelfont=bf,textfont=it]{caption}
\graphicspath{{Figures/}} 

\usepackage{geometry}
\usepackage{setspace}
\usepackage{multicol}
\usepackage{enumitem}
\usepackage[section]{placeins} 
\usepackage{tikz}
\usetikzlibrary{patterns, arrows.meta, positioning, intersections}

\usepackage{microtype}  

\usepackage{cite}
\usepackage{xcolor}
\usepackage[colorlinks=true,
linkcolor=blue,
citecolor=blue,
urlcolor=blue]{hyperref}
\usepackage{cleveref}   

\hypersetup{
	pdftitle={Electroosmotic lubrication flow in constricted microchannels with a compliant wall and DLVO interactions},
	pdfauthor={Subhajyoti Sahoo and Ameeya Kumar Nayak},
	pdfsubject={Electroosmotic lubrication in compliant microchannels},
	pdfkeywords={Electroosmosis, Compliant microchannels, DLVO, Elastohydrodynamics}
}

\usepackage{authblk}

\title{Electroosmotic lubrication in constricted microchannels with a compliant wall and DLVO disjoining pressure}
\author[1]{Subhajyoti Sahoo}
\author[1]{Ameeya Kumar Nayak\thanks{Corresponding author: \href{mailto:ameeya.nayak@ma.iitr.ac.in}{ameeya.nayak@ma.iitr.ac.in}}}
\affil[1]{Department of Mathematics, Indian Institute of Technology Roorkee, Roorkee, India}
\date{}

\begin{document}
	\maketitle
	
	\begin{abstract}
        We develop a nonlinear model for electroosmotic transport through a constricted microchannel with a rigid curved upper wall and a compliant lower wall. The framework couples Helmholtz--Smoluchowski slip under a globally constrained electric field, lubrication hydrodynamics, quasistatic Kirchhoff--Love wall bending, and Derjaguin--Landau--Verwey--Overbeek disjoining pressure. The response is governed by wall stiffness, geometric curvature, surface conduction, and the strengths of the repulsive and attractive intermolecular interactions. Asymptotic analysis and fully coupled Chebyshev-collocation computations reveal three regimes: a stiff-wall regime with negligible deformation, a compliance-limited regime in which localized throat narrowing suppresses throughput, and a small-gap saturation regime in which further thinning becomes progressively slower because of the coupled elastic, hydrodynamic, and disjoining-pressure response. In the strong-constriction limit, geometric focusing amplifies the electric field and produces a sublinear increase of throughput with curvature, whereas in the localized-throat limit the deformation and compliance-induced flux reduction are organized by the effective stiffness $\mathcal{B}C^2$. The results provide compact, design-oriented scaling laws for compliant electroosmotic constrictions and clarify how stiffness, curvature, surface conduction, and minimum-gap interactions regulate transport in soft microfluidic, biosensing, drug-delivery, and iontronic systems.
    \end{abstract}
	

\section{Introduction}
\label{sec:introduction}

Electroosmotic flow (EOF) has emerged as a premier mechanism for driving charged fluids and ions in micro- and nanofluidic systems\cite{saville1977electrokinetic, ren2001interfacial, probstein2005physicochemical, hunter2013zeta}, governed by the interaction of an applied electric field with mobile counterions within the electrical double layers (EDLs) adjacent to solid-liquid interfaces. In the thin-EDL limit, the fluid flow in the bulk region shows a plug-like profile with interfacial slip governed by the classical Helmholtz--Smoluchowski (HS) relation. Beyond this classical regime, induced charge electroosmosis and related nonlinear electrokinetic effects have been shown to modify the near wall slip and flow structure under strong applied fields \cite{schnitzer2012induced}. In this work we restrict attention to thin nonoverlapping double layers and quasi-Ohmic conduction, and we neglect induced charge electroosmosis, concentration polarization, and electrokinetic cross coupling. This framework underpins a wide range of applications in science and engineering, such as biosensing, chemical separations, drug delivery, and iontronic circuitry and electroosmotic diode/nanovalve devices \cite{wong2004electrokinetics,fiechtner2003faceted,gu2022electrokinetics,fernandez2021concentration,koyama2021electro}. Traditional microfluidic devices have primarily considered rigid channel walls with simple geometries, which enable analytical treatment of flow dynamics to optimize performance under different flow conditions. However, the shift toward soft and flexible materials in advanced microfluidics introduces new physics. Substrates such as polydimethylsiloxane (PDMS), hydrogels, and biological membranes possess elastic moduli orders of magnitude lower than those of conventional materials, allowing substantial wall deformation even under small flow induced pressure loads \cite{chakraborty2012fluid,yu2013deformable,ozsun2013non,mehboudi2018one,wang2019theory,guyard2022elastohydrodynamic,greidanus2022response}. These deformations alter the local channel geometry and pressure distribution resulting in a coupled fluid structure interaction (FSI) at low Reynolds numbers, where elastic and interfacial effects dominate the inertia \cite{essink2021regimes,karan2021generalization,roy2024fluid}.

FSI in electroosmotic systems exemplifies a broader class of problems where the compliant channel wall couples strongly to confined flows. Moreover, classical models of electroosmotic flow assume perfectly rigid channels, treating the geometry as a fixed parameter and ignoring solid mechanics entirely. However, introducing soft substrates brings wall deformation into the foreground and demands a framework to capture the bidirectional coupling. In this context, the appropriate mechanism for flow transport affected by variation in geometry and distribution of wall charge can be handled through the lubrication approximation theory \cite{ghosal2002lubrication}. Classical studies on EOF dealing with rigid surfaces primarily emphasized how the lubrication forces influence the colloidal mobility and Brownian dynamics. Recent work has analysed the stability and tunability of EOF in complex rigid wall configurations such as rotating microchannels and charge modulated periodic topographies \cite{shit2024stability,goyal2024generalizing}. However, when elasticity is introduced into the channel walls, qualitatively new behaviors emerge \cite{skotheim2005soft,skotheim2004dynamics,skotheim2004soft,salez2015elastohydrodynamics,bertin2022soft}. Elastic substrates give rise to elastohydrodynamic (EHD) coupling, wherein the pressure generated by the flow deforms the walls, and this deformation, in turn, feeds back on the flow field \cite{rallabandi2024fluid,bharti2024singular,paludan2024elastohydrodynamic,fares2024observation}. Such couplings have also been observed in surface forces apparatus (SFA) and atomic force microscopy (AFM), and exploited to extract the mechanical and rheological properties of compliant materials \cite{wang2015out,davies2018elastohydrodynamic}. At the level of scalar and charge transport, asymptotic analyses of dispersion in periodically modulated channels and multispecies electrolyte solutions further reveal how the microscale geometry and wall deformation interact to govern large-scale spreading \cite{ding2023shear,ding2025long}. These studies highlight the ubiquity of deformable interfaces in systems such as lubricated bearings, snow friction, and red blood cell motion through capillaries \cite{essink2021regimes}.

Deformation in compliant microchannels is studied by several authors in the context of pressure-driven flows, where hydrodynamic forces within the fluid act directly on the channel walls \cite{wang2022reduced,martinez2020start,inamdar2020unsteady}. In general, the pressure field associated with confined flow exerts a normal stress that deflects the channel walls, while viscous shear has less impact compared to tangential loads. The resulting deformation is resisted by the elastic restoring forces of the substrate, which are determined by the material properties such as Young’s modulus of the wall, Poisson ratio, and wall thickness. The coupling effect between the fluid stresses and solid elasticity gives rise to classical EHD interactions, in which the flow variation changes the channel geometry and alters hydraulic resistance \cite{ramos2021fluid,boyko2022flow}. Gervais et al.~\cite{gervais2006flow} introduced a compliance parameter to quantify this coupling effect and demonstrated that the characteristic wall deflection can be approximated as $\delta \sim p\,w^{4}/(E\,h^{3})$, where $p$ is the characteristic hydrodynamic pressure, $w$ is the channel width, $E$ is the Young's modulus of the substrate, and $h$ is the wall thickness. This scaling shows that PDMS microchannels can undergo a substantial deformation even at modest pressures, leading to a nonlinear increase in hydraulic resistance. 

Further theoretical advances have expanded this framework to capture how soft channels exhibit altered velocity profiles and develop mechanical instabilities, such as flow-induced collapse or valve-like closure under extreme confinement \cite{wang2019theory, wang2022reduced, boyko2022flow}. In addition to such quasi-static deformations, compliant microchannels can also exhibit dynamic interfacial phenomena, for example, intermittent air invasion driven by pervaporation \cite{keiser20222intermittent}. Many recent studies have considered this theory and extended it to viscoelastic and non-Newtonian fluids, highlighting the complex rheology coupling with the compliance parameter to reshape flow resistance \cite{boyko2023non,boyko2025interplay,chun2024flow}. This classical understanding of EHD theory provides the foundation for the present study and is further extended to involve the effect of electrokinetic phenomena to describe soft electrohydrodynamics coupling.

Fluid--structure interactions have been extensively studied for pressure-driven flows in soft microchannels; however, the corresponding coupled dynamics in electroosmotic transport remain largely unexplored. The classical electroosmotic theories assume rigid walls, and therefore do not capture the two-way coupling between electroosmotic driving, lubrication pressure, and wall deformation \cite{chakraborty2010influence, matse2018counterion,green2022effects, mandal2015electro, pal2024electroosmotic}. A few recent studies have started to explore electroosmotic or diffusioosmotic flow in compliant microchannels by demonstrating the electrokinetically augmented load bearing capacity and fluid structure instabilities in soft micro confinements \cite{mukherjee2022electrokinetically, boyko2020interfacial, bhaskaran2024elasto, maroundik2025diffusioosmotic}. Nevertheless, most of the studies in compliant microchannels have focused on mechanically forced or pressure-driven flows \cite{gervais2006flow,boyko2022flow,ramos2021fluid}, where electrokinetically induced deformation is often neglected to avoid simulation complexities. Even when compliance and electroosmosis are considered together, numerical treatments are considered for small deflections, linear elasticity, or idealized geometries, thereby missing the nonlinear feedbacks that really influence the governing phenomena associated with realistic soft microfluidic systems \cite{boyko2023non,chun2025experimental,boyko2025interplay}. Furthermore, intermolecular interactions such as van der Waals forces and electrostatic repulsion are rarely integrated with electrokinetic slip, despite their significant role when the gap becomes small while remaining in the thin nonoverlapping double layer regime \cite{rallabandi2024fluid,bharti2024singular,yaros2003evaluation,norouzisadeh2024a,mcnamee2019Effect}. These limitations hinder predictive modelling, obscure opportunities to deliberately harness wall compliance as a functional design parameter \cite{wang2019theory,martinez2020start,inamdar2020unsteady}, and reduce the reliability of performance in iontronic, biosensing, and drug delivery devices. Related analyses have also examined gaseous slip flow through shallow deformable microchannels, demonstrating that similar soft-lubrication mechanisms arise in rarefied gas regimes \cite{rodriguez2025gaseous}. Bridging these gaps is essential for advancing the theory of soft electroosmotic transport and can enable next-generation microfluidic systems that operate under coupled electrokinetic phenomena with wall deformable conditions.

In this paper, a comprehensive EOF model is developed in compliant microchannels involving the electrohydrodynamic theory with wall deformation. The mathematical formulation integrates the geometry-dependent electrokinetic slip, lubrication pressure in variable gaps, elastic wall mechanics, and intermolecular interactions to capture the nonlinear feedback that arises under small gaps while remaining in the thin nonoverlapping double layer regime. The incorporation of short-range intermolecular interactions through a Derjaguin-Landau-Verwey-Overbeek (DLVO) model is a key feature of this study; however, the primary novelty lies in the comprehensive, bidirectional coupling of these previously isolated physical domains. Specifically, this article introduces an augmented pressure formulation that couples hydrodynamic pressure, DLVO disjoining pressure, and Kirchhoff--Love plate deformation under an electroosmotic driving force. This is combined with a newly derived global voltage constraint that demonstrates how wall compliance and surface conduction dynamically alter electric field focusing through a harmonic mean conductance factor. Through this unified framework, we mathematically classify three distinct emergent transport regimes: a stiff-wall regime, a compliance-limited regime, and a small-gap saturation regime.Furthermore, the analysis provides novel, compact asymptotic scaling relations that explicitly predict maximum deflection and throughput variation in the strongly constricted, narrow-gap limit. These findings can be helpful to establish fundamental design principles for micro and nanofluidic devices and can also support the use of wall compliance as a functional control parameter in applications ranging from biosensing and analytical separations to drug delivery and soft iontronic systems. The framework also provides a foundation for experimental validation and extension to viscoelastic or charge-regulated systems.

The manuscript is organized as follows. Section \ref{sec:theory} introduces the theoretical framework for electrokinetics, hydrodynamics, elasticity, and intermolecular components of the fluid flow model, along with the nondimensional governing parameters. Section \ref{sec:asymptotics} presents the asymptotic analysis of the rigid-channel and small-deformation limits, providing the physical interpretation of the flow governing equations and balances. Section \ref{sec:numerics} describes the numerical methodology and validation. Section \ref{sec:results} discusses the emergent transport regimes, identifies the scaling relations that collapse the dynamics, and compares them against numerical results. Finally, Section \ref{sec:conclusion} summarizes the main findings, discusses their implications for soft electrokinetic systems, and highlights opportunities for future work in biosensing, drug delivery, and iontronic applications.

\section{Theoretical framework}
\label{sec:theory}

\begin{figure}[htbp]
    \centering
\begin{tikzpicture}[xscale=2.5,yscale=1.5,>=latex,font=\normalsize]
	
	\fill[gray!20] (0,-1.2) -- (5,-1.2) -- (5,0)
	.. controls (4.0,0) and (3.4,0.35) .. (2.5,0.35)
	.. controls (1.6,0.35) and (1.0,0) .. (0,0) -- cycle;
	
	\draw[ultra thick] (0,-1.2) -- (5,-1.2);
	\node[below] at (2.5,-1.2) {Compliant substrate};
	
	\draw[thick,->] (0,0) -- (5.4,0) node[right] {$x$};
	\draw[thick,->] (0,-1.2) -- (0,2.8) node[above] {$y$};
	
	\fill[pattern=north east lines,pattern color=black!60]
	(0,2.5) parabola bend (2.5,1.0) (5,2.5) -- (5,2.8) -- (0,2.8) -- cycle;
	\draw[ultra thick,name path=top_wall] (0,2.5) parabola bend (2.5,1.0) (5,2.5);
	
	\draw[thin,name path=undeformed_wall] (0,0) -- (5,0);
	
	\draw[thick,dashed,name path=deformed_wall]
	(0,0)
	.. controls (1.0,0) and (1.6,0.35) .. (2.5,0.35)
	.. controls (3.4,0.35) and (4.0,0) .. (5,0);
	
	\draw[->,thick] (0.35,1.2) -- (1.15,1.2) node[midway,above] {$E_x$};
	
	\draw[<->,thin] (0.4,0) -- (0.4,-1.2) node[midway,right] {$h_s$};
	
	\path[name path=vert_h0] (1.2,-1.0) -- (1.2,3.0);
	\draw[<->,thin,
	name intersections={of=undeformed_wall and vert_h0,by=h0_bottom},
	name intersections={of=top_wall and vert_h0,by=h0_top}]
	(h0_bottom) -- (h0_top) node[midway,right] {$h_0(x)$};
	
	\path[name path=vert_h] (3.8,-1.0) -- (3.8,3.0);
	\draw[<->,thin,
	name intersections={of=deformed_wall and vert_h,by=h_bottom},
	name intersections={of=top_wall and vert_h,by=h_top}]
	(h_bottom) -- (h_top) node[midway,right] {$h(x)=h_0(x)-\delta_s(x)$};
	
	\path[name path=vert_delta] (2.75,-1.0) -- (2.75,3.0);
	\draw[<->,thick,
	name intersections={of=undeformed_wall and vert_delta,by=delta_bottom},
	name intersections={of=deformed_wall and vert_delta,by=delta_top}]
	(delta_bottom) -- (delta_top) node[midway,right] {$\delta_s(x)$};
	
	\node[above] at (2.5,0.42) {$x=X_c$};
	
	\node[
	draw,
	rounded corners,
	fill=white,
	align=left,
	font=\small,
	inner sep=4pt
	] at (4.05,1.95)
	{$\delta_s>0$: upward / narrowing\\
		$\delta_s<0$: downward / opening};
	
\end{tikzpicture}
\caption{Electroosmotic lubrication in a constricted microchannel with a rigid curved upper wall and a compliant lower wall. A prescribed potential drop $\Delta\phi$ produces an axial electric field $E_x(x)$ in a symmetric $z{:}z$ electrolyte. Under quasi-one-dimensional conduction the field intensifies where the gap is small, thereby strengthening the Helmholtz--Smoluchowski electroosmotic slip. The undeformed and deformed lower-wall shapes are shown schematically. The wall displacement $\delta_s(x)$ is defined positive upward into the fluid. In the schematic shown, the compliant wall is displaced upward into the fluid, so $\delta_s(x)>0$ and therefore $h(x)<h_0(x)$. Conversely, $\delta_s(x)<0$ corresponds to downward displacement and gap opening.}
\label{fig:schematic}
\end{figure}

We consider steady two-dimensional transport of a symmetric $z{:}z$ electrolyte with permittivity $\varepsilon$, dynamic viscosity $\mu$, and bulk conductivity $\sigma_b$ in a channel of length $L$. The axial coordinate is $x\in[0,L]$. All fluxes and electric currents are reported per unit spanwise width. The instantaneous gap between the rigid upper wall and the compliant lower wall is denoted by $h(x)$.

\subsection{Geometry and lubrication assumptions}
\label{subsec:geometry}

The rigid upper wall is modeled as a circular arc of radius $R$ with its apex at $x=X_c$.  In the lubrication limit, the circular arc is approximated to leading order by its parabolic Taylor expansion, yielding
\begin{equation}
h_0(x)=H_0+\frac{(x-X_c)^2}{2R},
\end{equation}
where $H_0$ is the minimum undeformed gap at the throat. The lower-wall displacement $\delta_s(x)$ is defined positive upward, that is, toward the fluid. With this convention, a positive displacement narrows the channel and a negative displacement widens it. The instantaneous gap is therefore
\begin{equation}
h(x)=h_0(x)-\delta_s(x)
=H_0+\frac{(x-X_c)^2}{2R}-\delta_s(x),
\label{eq:gap}
\end{equation}
and we require $h(x)>0$ for all $x$. In particular, since $h_0(X_c)=H_0$, the throat gap satisfies $h(X_c)=H_0-\delta_s(X_c)$. We assume a slender geometry ($H_0/L\ll 1$), small slopes ($|h_x|\ll 1$, where $h_x\equiv dh/dx$), and small deflections ($|\delta_s|/H_0\ll 1$). Normalizing the quadratic geometric variation $L^2/(2R)$ by the minimum gap $H_0$ yields the convenient curvature parameter
\begin{equation}
C=\frac{L^2}{2RH_0},
\end{equation}
which is $O(1)$ for moderate constrictions. In the sharp-throat limit $C \gg 1$ detailed in Section \ref{sec:asymptotics}, we additionally require $\sqrt{C}\,(H_0/L) \ll 1$ to ensure the lubrication approximation remains valid within the inner throat region.

\subsection{Electrokinetics and current conservation}
\label{subsec:electrokinetics}

Electroosmosis results from the action of an applied electric field on the net charge in the electrical double layer (EDL). Throughout this work we restrict attention to thin, nonoverlapping EDLs,
\begin{equation}
\kappa h(x)\gg 1
\qquad \text{for all } x\in[0,L],
\label{eq:thinEDL}
\end{equation}
where $\kappa^{-1}$ is the Debye length. Since the local gap $h(x)$ depends on the elastic deformation, we verify this assumption a posteriori by computing
\begin{equation}
(\kappa h)_{\min}\equiv \kappa \min_{x\in[0,L]} h(x)
\qquad
\big(\text{equivalently, } (\kappa^\ast h)_{\min}=\kappa^\ast \min_{x\in[0,1]} h(x)\big),
\label{eq:kappa_hmin_def}
\end{equation}
for each numerical simulation. For all cases shown in Figs.~3--7 we find $(\kappa^\ast h)_{\min}\ge 10$, confirming that the EDLs remain thin and nonoverlapping even at the most constricted location. In the thin-EDL limit~\eqref{eq:thinEDL}, the leading-order electrokinetic effect reduces to an effective slip velocity given by the Helmholtz--Smoluchowski relation \cite{probstein2005physicochemical},
\begin{equation}
u_{\mathrm{eo}}(x)=-\frac{\varepsilon \zeta}{\mu} E_x(x),
\label{eq:ueo}
\end{equation}
where $\zeta$ is the prescribed zeta potential and $E_x(x)$ is the axial electric field. The imposed potential drop $\Delta\phi=\phi(0)-\phi(L)$ specifies the axial forcing through the bulk electric field $E_x(x)=-\frac{d\phi}{dx}$. The zeta potential $\zeta$ is an interfacial potential difference between the wall and the local bulk fluid just outside the EDL at the same axial location. A uniform gauge shift in the bulk potential changes neither $\Delta\phi$ nor $\zeta$. Prescribing $\Delta\phi$ and $\zeta$ therefore does not overdetermine the electric potential in the present formulation. We take $\Delta\phi>0$ without loss of generality, so that $\Delta\phi=\int_0^L E_x(x)\,dx$. Assuming insulating walls and a slender geometry ($H_0/L\ll 1$), the two-dimensional Laplace equation for the bulk potential asymptotically reduces to quasi-one-dimensional conduction at leading order. A full two-dimensional solution would capture only $O(H_0^2/L^2)$ corrections associated with streamline curvature and transverse field variations. The leading-order axial electric current per unit width is therefore conserved. Including bulk conduction through the gap and an identical surface conductance $K_s$ on each wall, we model the current as
\begin{equation}
I=\big(\sigma_b h(x)+2K_s\big)\,E_x(x),
\label{eq:ohm}
\end{equation}
where $I$ is independent of $x$. This closure neglects electrokinetic cross-coupling effects such as streaming current and concentration polarization. It is appropriate when Ohmic conduction dominates under the operating conditions considered. The imposed potential drop then satisfies
\begin{equation}
\Delta\phi
=I\int_0^L\frac{dx}{\sigma_b h(x)+2K_s}.
\label{eq:voltage_constraint}
\end{equation}
We characterize surface conduction through the Dukhin number $Du=\frac{K_s}{\sigma_b H_0}$.

\subsection{Lubrication hydrodynamics and disjoining pressure}
\label{subsec:hydro}

At sufficiently small gaps, the normal traction acting on the compliant wall includes both the hydrodynamic pressure and intermolecular surface forces. We represent the latter through a disjoining pressure $\Pi(h)$ and define the augmented pressure
\begin{equation}
P(x)=p(x)+\Pi\!\big(h(x)\big),
\label{eq:aug_pressure}
\end{equation}
where $p(x)$ is the hydrodynamic pressure. Using the nonretarded planar Derjaguin approximation \cite{israelachvili2011intermolecular}, we adopt a standard DLVO form consisting of an electrostatic repulsion and a van der Waals attraction,
\begin{equation}
\Pi(h)=
64\,n_0 k_B T\,\tanh^2\!\left(\frac{ze\psi_0}{4k_B T}\right)e^{-\kappa h}
-\frac{A_H}{6\pi h^3},
\label{eq:dlvo}
\end{equation}
where $n_0$ is the bulk ion concentration, $\psi_0$ is the Stern-plane potential, and $A_H$ is the Hamaker constant. The exponential repulsive term is consistent with the nonoverlapping-EDL regime considered in Sec.~\ref{subsec:electrokinetics}, since all reported solutions satisfy $\kappa h_{\min}\gg 1$. Even when $\kappa h\gg 1$, the repulsive contribution can still affect the mechanics because its dimensional prefactor may be large enough for the product of that prefactor and the exponential factor to remain comparable to hydrodynamic or elastic stresses near the minimum gap. For $\kappa h=O(1)$ or smaller, or when $h$ approaches molecular scales, additional physics such as steric or hydration forces or charge regulation may be required; such regimes are not considered here.

The electrokinetic zeta potential $\zeta$ entering the slip condition and the Stern-plane potential $\psi_0$ entering the DLVO repulsion are treated as distinct prescribed parameters. The role of $\Pi(h)$ in the flow follows directly from the classical lubrication momentum balance and mass conservation laws \cite{leal2007advanced}. In this asymptotic framework, intermolecular interactions enter as an additional isotropic normal stress, so the leading-order axial Stokes balance becomes
\begin{equation}
0=-\frac{d}{dx}\Big(p(x)+\Pi\!\big(h(x)\big)\Big)+\mu \frac{\partial^2 u}{\partial y^2}
=-\frac{dP}{dx}+\mu \frac{\partial^2 u}{\partial y^2},
\label{eq:stokes_aug}
\end{equation}
where $u(x,y)$ is the axial velocity. Thus the disjoining pressure enters the fluid momentum balance only through the axial gradient of the augmented pressure $P(x)$, not as a tangential wall traction. A useful consistency check is the static limit: in the absence of electroosmotic driving, a quiescent fluid state ($u\equiv 0$) requires $dP/dx=0$. Hence any axial variation of $\Pi(h(x))$ induced by the slowly varying gap is exactly balanced by an opposing hydrodynamic pressure gradient $dp/dx$, thereby preventing any spurious tangential flow. Under lubrication theory, the depth-averaged volumetric flux per unit width is the sum of a Poiseuille contribution driven by the augmented pressure gradient and an electroosmotic plug contribution. For identical electrokinetic properties on both walls,
\begin{equation}
q(x)=-\frac{h^3(x)}{12\mu}\frac{dP}{dx}+h(x)\,u_{\mathrm{eo}}(x),
\label{eq:flux}
\end{equation}
where $u_{\mathrm{eo}}(x)$ is given by Eq.~\eqref{eq:ueo}. Steady incompressibility implies $dq/dx=0$, and hence $q(x)=Q$ is constant.

\subsection{Wall mechanics and pressure reference}
\label{subsec:mechanics}

The compliant lower wall is modeled as a clamped Kirchhoff--Love plate strip of thickness $h_s$ and flexural rigidity \cite{timoshenko1959theory}
\begin{equation}
D=\frac{E_Y h_s^3}{12(1-\nu^2)},
\label{eq:rigidity}
\end{equation}
where $E_Y$ and $\nu$ are the Young's modulus and Poisson ratio. In the lubrication limit, normal tractions dominate tangential viscous stresses, so the fluid--structure coupling is primarily through the normal load $P(x)$. Unlike rigid-wall lubrication, the plate response depends on the absolute normal traction level. Following standard hydrodynamic conventions \cite{leal2007advanced}, we fix the gauge by taking the external backside pressure as the reference and imposing equal reservoir gauge pressures
\begin{equation}
p(0)=p(L)=0.
\label{eq:pressureBC}
\end{equation}
With $\delta_s$ defined positive upward into the fluid, a positive augmented pressure acts downward on the lower wall and therefore gives
\begin{equation}
D\frac{d^4\delta_s}{dx^4}=-P(x)=-\big(p(x)+\Pi(h)\big).
\label{eq:beam}
\end{equation}
Here and throughout, the compact notation $\delta_s^{(4)}(x)\equiv d^4\delta_s/dx^4$ denotes the fourth derivative with respect to $x$. With this convention, $P>0$ tends to drive the lower wall downward, corresponding to $\delta_s<0$ in the loaded region, whereas upward narrowing of the gap requires $P<0$. In particular, positive hydrodynamic pressure $p$ by itself tends to deflect the lower wall downward and hence to increase the local gap. The wall is clamped at both ends,
\begin{equation}
\delta_s(0)=\delta_s(L)=0,
\qquad
\delta_s'(0)=\delta_s'(L)=0.
\label{eq:clamped_dim}
\end{equation}
If membrane tension is relevant, an additional term $T\,\delta_s''$ may be included. Here we focus on bending-dominated response.

\subsection{Nondimensionalization and coupled equations}
\label{subsec:nondim}

We nondimensionalize according to
\[
x=L\tilde{x},\qquad
h=H_0\tilde{h},\qquad
\delta_s=H_0\tilde{\delta}_s,\qquad
q=U_\star H_0\tilde{q},\qquad
p=p_\star\tilde{p},
\]
and then drop the tildes for notational simplicity. We use the electroosmotic velocity scale $U_\star=\frac{\varepsilon |\zeta| \Delta\phi}{\mu L}$, and the lubrication pressure scale $p_\star=\frac{\mu U_\star L}{H_0^2}$, where $\Delta\phi=\phi(0)-\phi(L)>0$ is the imposed axial potential drop. The dimensionless gap is
\begin{equation}
h(x)=1+C(x-x_c)^2-\delta_s(x),
\qquad
x_c=\frac{X_c}{L}.
\label{eq:nd_gap}
\end{equation}
The key dimensionless parameters are
\[
\kappa^\ast=\kappa H_0,\qquad
Du=\frac{K_s}{\sigma_b H_0},\qquad
\Pi_{\mathrm{el}}=\frac{64n_0k_BT\tanh^2\!\left(\frac{ze\psi_0}{4k_BT}\right)}{p_\star},\qquad
\Pi_{\mathrm{vdW}}=\frac{A_H/(6\pi H_0^3)}{p_\star},\qquad
\mathcal{B}=\frac{D H_0^3}{\mu U_\star L^5},
\]
where $\mathcal{B}$ is the dimensionless bending stiffness. The lubrication approximation also assumes a small fluid Reynolds number, $Re_f=\frac{\rho U_\star H_0}{\mu}\ll 1$. We define the dimensionless electric field by $\mathcal{E}(x)=\frac{E_x L}{\Delta\phi}$. From Eqs.~\eqref{eq:ohm} and \eqref{eq:voltage_constraint} we obtain
\begin{equation}
\mathcal{E}(x)=\frac{\alpha}{h(x)+2Du},
\qquad
\alpha=\left(\int_0^1 \frac{dx}{h(x)+2Du}\right)^{-1}.
\label{eq:nd_field}
\end{equation}
Let $s_\zeta=\frac{\zeta}{|\zeta|}=\pm 1$. Then Eq.~\eqref{eq:ueo} gives $\frac{u_{\mathrm{eo}}}{U_\star}=-s_\zeta \mathcal{E}(x)$, so that a negatively charged wall ($\zeta<0$) corresponds to $s_\zeta=-1$ and hence to positive electroosmotic slip in the positive $x$ direction when $\mathcal{E}>0$. The dimensionless augmented pressure is
\begin{equation}
P=p+\Pi_{\mathrm{el}}e^{-\kappa^\ast h}-\frac{\Pi_{\mathrm{vdW}}}{h^3}.
\label{eq:nd_augP}
\end{equation}
The dimensionless flux law corresponding to Eq.~\eqref{eq:flux} becomes
\begin{equation}
Q=-\frac{h^3}{12}\frac{dP}{dx}-s_\zeta \frac{\alpha h}{h+2Du}.
\label{eq:nd_flux}
\end{equation}
The plate equation~\eqref{eq:beam} becomes
\begin{equation}
\mathcal{B}\frac{d^4\delta_s}{dx^4}=-P,
\label{eq:nd_beam}
\end{equation}
since $\mathcal{B}=D H_0^3/(\mu U_\star L^5)=D H_0/(p_\star L^4)$. Thus the sign convention remains unchanged after nondimensionalization: positive $P$ drives the lower wall downward, whereas gap narrowing requires negative $P$. The clamped boundary conditions are $\delta_s(0)=\delta_s(1)=0$, $\delta_s'(0)=\delta_s'(1)=0$, and the pressure gauge is
$p(0)=p(1)=0$. To obtain an explicit solvability condition for $Q$, we divide Eq.~\eqref{eq:nd_flux} by $h^3$ and integrate over $x\in[0,1]$. Using $p(0)=p(1)=0$, we obtain
\begin{equation}
Q=
\frac{
-s_\zeta \alpha \displaystyle\int_0^1 \frac{dx}{h^2(x)\big(h(x)+2Du\big)}
-\frac{1}{12}\left[
\Pi_{\mathrm{el}}\Big(e^{-\kappa^\ast h(1)}-e^{-\kappa^\ast h(0)}\Big)
-\Pi_{\mathrm{vdW}}\left(\frac{1}{h^3(1)}-\frac{1}{h^3(0)}\right)
\right]
}{
\displaystyle\int_0^1 h^{-3}(x)\,dx
}.
\label{eq:Q_solv}
\end{equation}
For equal end gaps, the boundary term vanishes. In the Ohmic limit $Du\to 0$, the integral ratio tends to unity and the electroosmotic contribution reduces to $Q=-s_\zeta \alpha$. Equations~\eqref{eq:nd_gap}--\eqref{eq:nd_beam}, together with the voltage constraint~\eqref{eq:nd_field} and the stated boundary conditions, define the coupled electroosmotic lubrication and wall-bending problem with disjoining pressure in the thin-EDL, quasi-Ohmic regime.

\section{Asymptotic analysis}
\label{sec:asymptotics}

Based on the dimensionless model in Sec.~\ref{sec:theory}, we analyze limiting regimes in which simplified balances reveal the dominant mechanisms that govern augmented pressure, wall deformation, and throughput. We use the dimensionless gap $h(x)=1+C(x-x_c)^2-\delta_s(x)$, the augmented pressure $P(x)=p(x)+\Pi_{\mathrm{el}}e^{-\kappa^\ast h(x)}-\frac{\Pi_{\mathrm{vdW}}}{h^3(x)}$, the voltage constraint $\alpha=\left(\int_0^1\frac{dx}{h(x)+2Du}\right)^{-1}$, and the flux law
\begin{equation}
Q=-\frac{h^3}{12}\frac{dP}{dx}-s_\zeta\frac{\alpha h}{h+2Du}.
\label{eq:asy_flux}
\end{equation}
Here $s_\zeta=\zeta/|\zeta|=\pm 1$ and the reference velocity scale is $U_\star=\varepsilon|\zeta|\Delta\phi/(\mu L)$. With $\delta_s$ defined positive upward into the fluid, a positive compressive normal traction pushes the wall downward. The bending equilibrium is
\begin{equation}
P(x)=-\mathcal{B}\frac{d^4\delta_s}{dx^4},
\label{eq:asy_plate}
\end{equation}
where $\mathcal{B}$, the dimensionless bending stiffness, is introduced in Sec.~\ref{subsec:nondim}. The results below provide physical interpretation of the coupled system and benchmarks for the numerical solutions.

\subsection{Weak compliance limit with a stiff wall}
\label{sec:stiff}

We consider $\mathcal{B}\gg 1$, corresponding to a stiff wall with small deflection. Define $\epsilon=\mathcal{B}^{-1}\ll 1$ and expand
\[
p=p_0+\epsilon p_1+\cdots,\qquad
\delta_s=\epsilon\delta_1+\cdots,\qquad
h=h_0-\epsilon\delta_1+\cdots,\qquad
Q=Q_0+\epsilon Q_1+\cdots,
\]
with the undeformed gap $h_0(x)=1+C(x-x_c)^2$. In this limit, geometry is fixed to leading order, and wall deformation enters as a perturbation. We restrict our analytical expansion to $O(\epsilon)$ to cleanly isolate the dominant linear coupling mechanisms, specifically how small deflections modify hydraulic resistance and electrical conductance. Higher-order nonlinear feedbacks are natively captured by the fully coupled numerical framework developed in Sec. 4. The leading order pressure and disjoining stresses provide the forcing for the first order deflection. The first order correction captures how small changes of the gap modify both hydraulic resistance and electrical conductance.

At $O(1)$, Eq.~\eqref{eq:asy_plate} gives
\begin{equation}
\delta_1^{(4)}(x)=-P_0(x),
\qquad
P_0(x)=p_0(x)+\Pi_{\mathrm{el}}e^{-\kappa^\ast h_0(x)}-\frac{\Pi_{\mathrm{vdW}}}{h_0^3(x)}.
\label{eq:stiff_P0}
\end{equation}
Here and throughout, the notation $f^{(4)}$ denotes the fourth derivative with respect to the independent variable. In particular, $\delta_1^{(4)}(x)\equiv d^4\delta_1/dx^4$. This fourth derivative arises from the Kirchhoff--Love bending operator used for the compliant wall. At the same order, substituting $h=h_0$ and $P=P_0$ into Eq.~\eqref{eq:asy_flux} yields
\begin{equation}
Q_0=-\frac{h_0^3}{12}\frac{dP_0}{dx}-s_\zeta\frac{\alpha_0 h_0}{h_0+2Du},
\qquad
\alpha_0=\left(\int_0^1\frac{dx}{h_0(x)+2Du}\right)^{-1}.
\label{eq:stiff_Q0}
\end{equation}
In the Ohmic limit $Du\to 0$, the electroosmotic contribution simplifies to $-s_\zeta\alpha_0$. This spatially uniform electroosmotic flux reflects the underlying physics that the local slip velocity $u_{eo}(x)$ increases inversely with the narrowing gap to satisfy flow continuity and current conservation. The factor $\alpha_0$ is a harmonic mean conductance that weights the constricted region most strongly. The disjoining contribution linearized about $h_0$ is
\begin{equation}
\Pi_{\mathrm{el}}e^{-\kappa^\ast h}-\frac{\Pi_{\mathrm{vdW}}}{h^3}
=
\left(\Pi_{\mathrm{el}}e^{-\kappa^\ast h_0}-\frac{\Pi_{\mathrm{vdW}}}{h_0^3}\right)
+\epsilon\delta_1\left(\kappa^\ast\Pi_{\mathrm{el}}e^{-\kappa^\ast h_0}-\frac{3\Pi_{\mathrm{vdW}}}{h_0^4}\right)
+O(\epsilon^2).
\label{eq:stiff_DLVOexp}
\end{equation}
This defines the incremental disjoining stiffness
\[
K_{\mathrm{DLVO}}(x)\equiv -\frac{d\Pi}{dh}\Big|_{h=h_0}
=\kappa^\ast\Pi_{\mathrm{el}}e^{-\kappa^\ast h_0(x)}-\frac{3\Pi_{\mathrm{vdW}}}{h_0^4(x)}.
\]
Positive $K_{\mathrm{DLVO}}$ corresponds to repulsion dominated stiffening, while negative $K_{\mathrm{DLVO}}$ corresponds to attraction dominated softening. The voltage factor expands as
\begin{equation}
\alpha=\alpha_0+\epsilon\alpha_1+\cdots,
\qquad
\alpha_1=-\alpha_0^2\int_0^1\frac{\delta_1(x)}{(h_0(x)+2Du)^2}\,dx.
\label{eq:stiff_alpha1}
\end{equation}
This correction shows that a small deflection changes electrical resistance through the factor $(h+2Du)^{-1}$ and the effect is weighted toward the throat. Writing $P=P_0+\epsilon P_1+\cdots$ with $P_1(x)=p_1(x)+\delta_1(x)K_{\mathrm{DLVO}}(x)$, the $O(\epsilon)$ expansion of Eq.~\eqref{eq:asy_flux} yields
\begin{align}
q_1(x)
&=
-\frac{h_0^3}{12}\frac{dP_1}{dx}
+\frac{h_0^2\delta_1}{4}\frac{dP_0}{dx}
+s_\zeta\left[
\frac{2\alpha_0Du\delta_1}{(h_0+2Du)^2}
-\frac{h_0}{h_0+2Du}\alpha_1
\right].
\label{eq:stiff_q1}
\end{align}
Since the flux is spatially uniform, $Q_1$ follows from a solvability condition enforcing constant total flux, for example by averaging $q_1$ over $x\in[0,1]$. The deflection satisfies clamped conditions $\delta_1(0)=\delta_1(1)=\delta_1'(0)=\delta_1'(1)=0$.

\subsection{Strong constriction limit with $C\gg 1$}
\label{sec:strongC}

For a sharp constriction with $C\gg 1$, the pressure and deformation localize near the throat. Introduce the stretched coordinate
\[
\xi=\sqrt{C}(x-x_c),
\qquad
dx=\frac{d\xi}{\sqrt{C}},
\qquad
h(\xi)=1+\xi^2-\delta(\xi),
\]
where $\delta(\xi)=\delta_s(x(\xi))$ in inner variables. Derivatives transform as $d/dx=\sqrt{C}\,d/d\xi$, hence $d^4/dx^4=C^2 d^4/d\xi^4$. We write $\delta^{(4)}(\xi)\equiv d^4\delta/d\xi^4$ for the fourth derivative with respect to $\xi$. We require $\sqrt{C}(H_0/L)\ll 1$ so that lubrication remains valid in the throat region.

Furthermore, achieving the $C \gg 1$ limit by decreasing the minimum gap $H_0$ is physically bounded by the thin-EDL requirement. For the Helmholtz--Smoluchowski slip formulation to remain valid, we strictly require $\kappa H_0 \gg 1$ so that the electrical double layers do not overlap at the throat. Thus, the strong constriction limit implicitly assumes that the electrolyte concentration is sufficiently high to maintain a Debye length that is much smaller than the narrowed gap. We also assume the throat is not close to the ends so that the inner region does not interact with boundary conditions at leading order. The voltage constraint weights small gaps most strongly, so electrical resistance is throat dominated when $C$ is large and $Du$ is not dominant. To obtain the leading scaling we neglect $\delta$ in the voltage integral. Extending the throat dominated integral to $\xi\in(-\infty,\infty)$ gives
\[
\alpha^{-1}=\int_0^1\frac{dx}{h(x)+2Du}
\sim \frac{1}{\sqrt{C}}\int_{-\infty}^{\infty}\frac{d\xi}{1+\xi^2+2Du}
=\frac{\pi}{\sqrt{C}\sqrt{1+2Du}},
\]
hence
\begin{equation}
\alpha\sim \frac{\sqrt{C}}{\pi}\sqrt{1+2Du}.
\label{eq:alpha_strongC}
\end{equation}
When $Du\gg C$, surface conduction dominates and the integrand is nearly uniform, so the throat no longer controls the voltage integral. For equal end gaps, the closure reduces to
\[
Q=-s_\zeta\alpha\frac{I_1}{I_2},
\qquad
I_1=\int_0^1\frac{dx}{h^2(x)(h(x)+2Du)},
\qquad
I_2=\int_0^1 h^{-3}(x)\,dx.
\]
Evaluating throat dominated integrals yields
\begin{equation}
Q\sim -s_\zeta\sqrt{C}K(Du),
\label{eq:Q_strongC}
\end{equation}
where
\begin{equation}
K(Du)=\frac{8}{3\pi^2}\sqrt{1+2Du}A(Du),
\qquad
A(Du)=\int_{-\infty}^{\infty}\frac{d\xi}{(1+\xi^2)^2(1+\xi^2+2Du)}.
\label{eq:K_def}
\end{equation}
The integral evaluates to
\begin{equation}
A(Du)=\frac{\pi}{4Du^2}\left(Du-1+\frac{1}{\sqrt{1+2Du}}\right),
\label{eq:A_closed}
\end{equation}
so that
\begin{equation}
K(Du)=\frac{2}{3\pi}\frac{\sqrt{1+2Du}(Du-1)+1}{Du^2},
\qquad
K(0)=\frac{1}{\pi}.
\label{eq:K_closed}
\end{equation}
For $Du\gg 1$ with $C\gg 1+2Du$, $K(Du)\sim (2\sqrt{2}/3\pi)Du^{-1/2}$. Physically, large surface conductance reduces the bulk current and weakens the effective electroosmotic mobility. In inner variables, Eq.~\eqref{eq:asy_plate} becomes
\begin{equation}
P(\xi)=-\mathcal{B}C^2\delta^{(4)}(\xi),
\label{eq:inner_plate}
\end{equation}
since $d^4/dx^4=C^2 d^4/d\xi^4$. A load of characteristic magnitude $P_{\max}$ varying over $\xi=O(1)$ therefore produces
\begin{equation}
|\delta|_{\max}\sim \frac{P_{\max}}{\mathcal{B}C^2}.
\label{eq:delta_strongC}
\end{equation}
The factor $C^2$ reflects that bending resistance increases rapidly when deformation is localized to a region of width $C^{-1/2}$.

\subsection{Thin gap scaling and DLVO crossovers}
\label{sec:thin_gap_scaling}

Let $h_{\min}$ denote the minimum throat gap, typically near $\xi=0$. A consistent disjoining forcing scale is
\[
F_{\mathrm{DLVO}}\sim \max\left(\Pi_{\mathrm{el}}e^{-\kappa^\ast h_{\min}},\frac{\Pi_{\mathrm{vdW}}}{h_{\min}^3}\right).
\]
Using Eq.~\eqref{eq:delta_strongC} gives
\begin{equation}
|\delta|_{\max}\sim \frac{F_{\mathrm{DLVO}}}{\mathcal{B}C^2}.
\label{eq:deltamax_scaling}
\end{equation}
The throughput change arises primarily through the global conductance factor $\alpha$, which scales as $\sqrt{C}$ in the throat dominated regime. Therefore,
\begin{equation}
|\Delta Q|\sim \sqrt{C}|\delta|_{\max}\sim \frac{F_{\mathrm{DLVO}}}{\mathcal{B}C^{3/2}}.
\label{eq:dQ_scaling}
\end{equation}
In the Ohmic limit $Du=0$, the ratio $I_1/I_2$ equals unity for any gap profile, so throughput changes are governed entirely by the conductance correction through $\alpha$. The dimensionless disjoining pressure is
\begin{equation}
\Pi(h)=\Pi_{\mathrm{el}}e^{-\kappa^\ast h}-\frac{\Pi_{\mathrm{vdW}}}{h^3}.
\label{eq:dlvo_nd}
\end{equation}
A crossover thickness $h_c$ at which repulsive and attractive contributions have comparable magnitude is defined by
\begin{equation}
\Pi_{\mathrm{el}}e^{-\kappa^\ast h_c}=\frac{\Pi_{\mathrm{vdW}}}{h_c^3}.
\label{eq:hc_general}
\end{equation}
For $\kappa^\ast h_c\gg 1$ the repulsive term is strongly screened and the interaction is attraction dominated. A formal estimate obtained by expanding $e^{-\kappa^\ast h}$ for $\kappa^\ast h\ll 1$ gives $h_c\sim(\Pi_{\mathrm{vdW}}/\Pi_{\mathrm{el}})^{1/3}$. This small gap expansion lies outside the thin nonoverlapping EDL regime assumed for both the slip model and the exponential DLVO form and it is used only as a qualitative scaling of the chosen constitutive law. The incremental disjoining stiffness is
\begin{equation}
K_{\mathrm{DLVO}}(h)\equiv -\frac{d\Pi}{dh}
=\kappa^\ast\Pi_{\mathrm{el}}e^{-\kappa^\ast h}-\frac{3\Pi_{\mathrm{vdW}}}{h^4}.
\label{eq:Kdlvo_general}
\end{equation}
A marginal thickness $h_m$ at which $K_{\mathrm{DLVO}}(h_m)=0$ satisfies
\begin{equation}
\kappa^\ast\Pi_{\mathrm{el}}e^{-\kappa^\ast h_m}=\frac{3\Pi_{\mathrm{vdW}}}{h_m^4}.
\label{eq:hm_general}
\end{equation}
A compact measure of repulsion versus attraction is
\begin{equation}
\Pi_{\mathrm{ratio}}(h)=\frac{\Pi_{\mathrm{el}}e^{-\kappa^\ast h}h^3}{\Pi_{\mathrm{vdW}}}.
\label{eq:Pi_ratio}
\end{equation}
Repulsion dominates for $\Pi_{\mathrm{ratio}}\gg 1$, attraction dominates for $\Pi_{\mathrm{ratio}}\ll 1$, and $\Pi_{\mathrm{ratio}}=O(1)$ indicates comparable magnitudes.

\section{Numerical methodology and validation}
\label{sec:numerics}

The coupled electroosmotic lubrication and wall bending model is defined by the gap relation in Eq.~\eqref{eq:nd_gap}, the voltage constraint in Eq.~\eqref{eq:nd_field}, the augmented pressure in Eq.~\eqref{eq:nd_augP}, the flux law in Eq.~\eqref{eq:nd_flux}, and the plate equilibrium in Eq.~\eqref{eq:nd_beam}, together with clamped boundary conditions and the pressure gauge condition. This system forms a stiff nonlinear boundary value problem. Stiffness arises from the fourth order bending operator, the exponential dependence of the repulsive disjoining term, and the algebraic singularity of the attractive term as $h\to 0$. We discretize the governing equations using Chebyshev spectral collocation and solve the resulting nonlinear algebraic system using a Newton--Raphson method with backtracking line search. Strongly nonlinear branches are obtained by continuation in selected parameters.

\subsection{Discretization and residual formulation}

The domain $x\in[0,1]$ is discretized on $N+1$ Chebyshev Gauss Lobatto nodes mapped to $[0,1]$
\begin{equation}
x_j=\frac12\left(1-\cos\left(\frac{j\pi}{N}\right)\right),\qquad j=0,\dots,N.
\label{eq:cheb_nodes}
\end{equation}
At the collocation nodes we approximate $\delta_s(x_j)\approx \delta_j$ and $p(x_j)\approx p_j$, and compute the gap from Eq.~\eqref{eq:nd_gap} as
\begin{equation}
h_j=1+C(x_j-x_c)^2-\delta_j.
\label{eq:h_nodes}
\end{equation}
Spatial derivatives are evaluated using Chebyshev differentiation matrices $\mathbf D$, $\mathbf D^2$, and $\mathbf D^4$ on $[0,1]$. Clenshaw--Curtis quadrature weights $w_j$ mapped to $[0,1]$ are used for spectrally accurate integration. The nodal augmented pressure is computed from Eq.~\eqref{eq:nd_augP} as
\begin{equation}
P_j=p_j+\Pi_{\mathrm{el}}e^{-\kappa^\ast h_j}-\frac{\Pi_{\mathrm{vdW}}}{h_j^3}.
\label{eq:P_nodes}
\end{equation}
The conductance factor $\alpha$ is recomputed from the discrete form of Eq.~\eqref{eq:nd_field}
\begin{equation}
\alpha=\left(\sum_{j=0}^{N}\frac{w_j}{h_j+2Du}\right)^{-1}.
\label{eq:alpha_disc}
\end{equation}
The unknown vector is $\mathbf u=(\boldsymbol\delta,\boldsymbol p,Q)$, where $\boldsymbol\delta$ and $\boldsymbol p$ collect nodal values and $Q$ is the constant flux. The residual includes the plate equilibrium at the collocation nodes, the flux law at the collocation nodes, and one scalar closure enforcing equal reservoir pressures. The plate residual enforces Eq.~\eqref{eq:nd_beam} in the form $P+\mathcal{B}\delta_s^{(4)}=0$
\begin{equation}
\mathcal R_1^{(j)}=P_j+\mathcal{B}\left[\mathbf D^4\boldsymbol\delta\right]_j.
\label{eq:R_plate}
\end{equation}
The flux residual enforces Eq.~\eqref{eq:nd_flux}
\begin{equation}
\mathcal R_2^{(j)}=
-\frac{h_j^3}{12}\left[\mathbf D\mathbf P\right]_j
-s_\zeta\frac{\alpha h_j}{h_j+2Du}
-Q,
\qquad j=0,\dots,N,
\label{eq:R_flux}
\end{equation}
where $s_\zeta=\zeta/|\zeta|=\pm 1$. Clamped wall conditions from Eq.~\eqref{eq:clamped_dim} are imposed by row replacement
\begin{equation}
\delta_0=0,\qquad \delta_N=0,\qquad
\left[\mathbf D\boldsymbol\delta\right]_0=0,\qquad
\left[\mathbf D\boldsymbol\delta\right]_N=0.
\label{eq:bc_clamped_disc}
\end{equation}
The pressure gauge condition $p(0)=0$ is imposed by replacing one equation with
\begin{equation}
p_0=0.
\label{eq:bc_gauge_disc}
\end{equation}
Equal reservoir pressures $p(1)=p(0)$ are enforced through the scalar closure
\begin{equation}
\mathcal R_3=p_N-p_0=0.
\label{eq:R_closure}
\end{equation}
Under the gauge in Eq.~\eqref{eq:bc_gauge_disc}, Eq.~\eqref{eq:R_closure} yields $p_N=0$ and determines the unknown flux $Q$ consistently with the continuous boundary condition.

\subsection{Nonlinear solution algorithm}

Let $\mathcal F(\mathbf u)$ denote the assembled residual built from Eq.~\eqref{eq:R_plate}, Eq.~\eqref{eq:R_flux}, and Eq.~\eqref{eq:R_closure}. At Newton iteration $n$ we solve
\begin{equation}
\mathcal J^{(n)}\Delta\mathbf u=-\mathcal F^{(n)},
\qquad
\mathbf u^{(n+1)}=\mathbf u^{(n)}+\lambda_{\mathrm{step}}\Delta\mathbf u,
\label{eq:newton_step}
\end{equation}
where $0<\lambda_{\mathrm{step}}\le 1$ is chosen by backtracking line search. We use an inexact Newton strategy in which $\alpha$ is recomputed from Eq.~\eqref{eq:alpha_disc} using the current gap values $h_j$ at each outer iteration and held fixed within the Jacobian. Including the sensitivity
\begin{equation}
\frac{\partial \alpha}{\partial h_j}=\alpha^2\frac{w_j}{(h_j+2Du)^2}
\label{eq:alpha_sens}
\end{equation}
in $\mathcal J$ yields indistinguishable solutions with modestly fewer iterations. To improve robustness when the Jacobian is ill conditioned at large $N$, we apply diagonal equilibration of rows and columns in the linear solve. To avoid transient van der Waals singularities during the line search, we enforce $h_j>h_{\mathrm{floor}}$ with $h_{\mathrm{floor}}=10^{-4}$. Converged solutions remain above this floor. Initial guesses are taken from the stiff-wall state with $\delta_s\equiv 0$, and strongly nonlinear cases are obtained by continuation in $\Pi_{\mathrm{vdW}}$, $\Pi_{\mathrm{el}}$, or $\mathcal{B}^{-1}$. Convergence is declared when
\begin{equation}
\|\mathcal F^{(n)}\|_\infty<10^{-8},
\qquad
\|\Delta\boldsymbol\delta\|_\infty<10^{-8},
\qquad
\|\Delta\boldsymbol p\|_\infty<10^{-8},
\qquad
|\Delta Q|<10^{-10}.
\label{eq:conv_criteria}
\end{equation}
These tolerances are typically reached within 10 to 20 iterations. Convergence is quadratic in weakly nonlinear regimes and remains robust under strong disjoining forcing when under relaxation is required.

\subsection{Validation and grid convergence}

\begin{figure}[t]
\centering
\begin{subfigure}{0.32\textwidth}
\centering
\includegraphics[height=5cm,width=\linewidth]{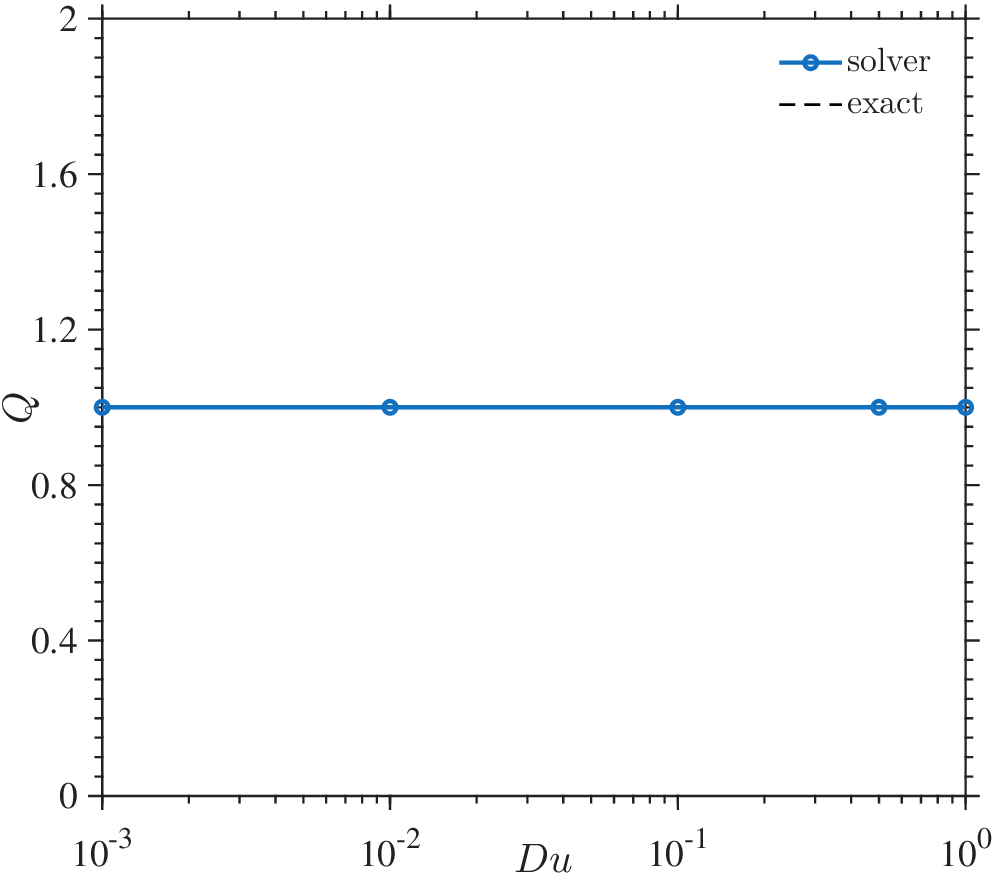}
\caption{}
\label{fig:validation_a}
\end{subfigure}
\hfill
\begin{subfigure}{0.32\textwidth}
\centering
\includegraphics[height=5cm,width=\linewidth]{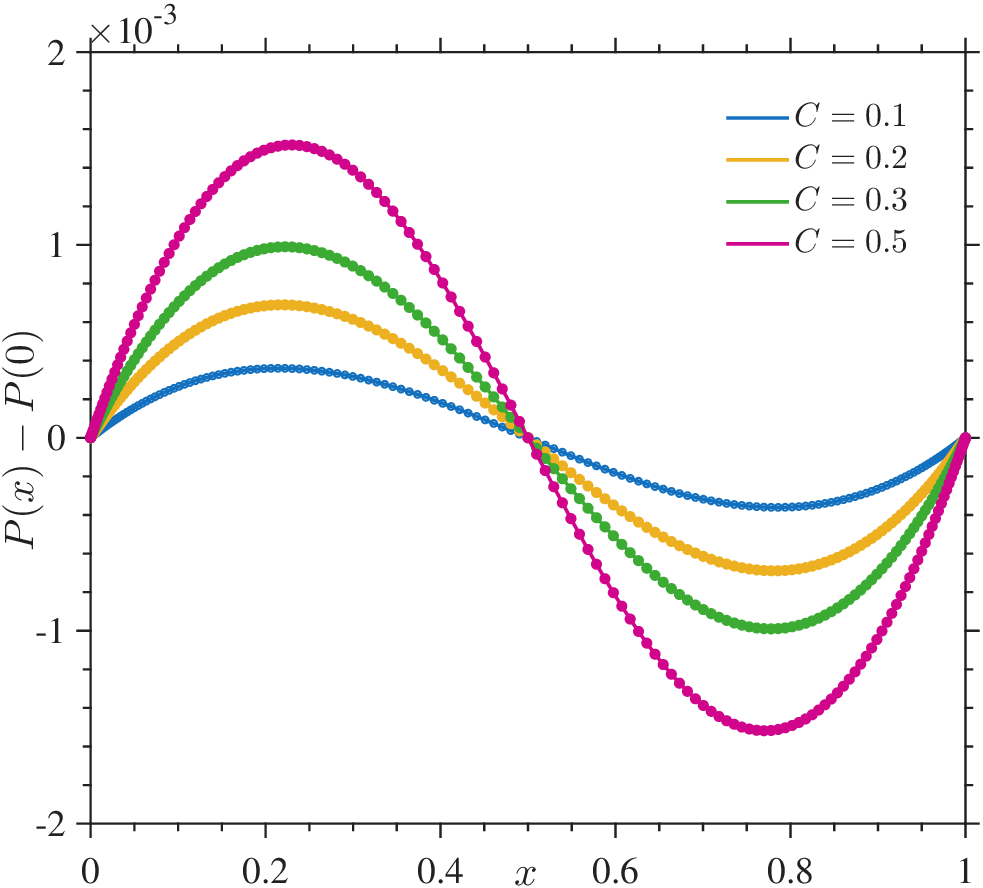}
\caption{}
\label{fig:validation_b}
\end{subfigure}
\hfill
\begin{subfigure}{0.32\textwidth}
\centering
\includegraphics[height=5cm,width=\linewidth]{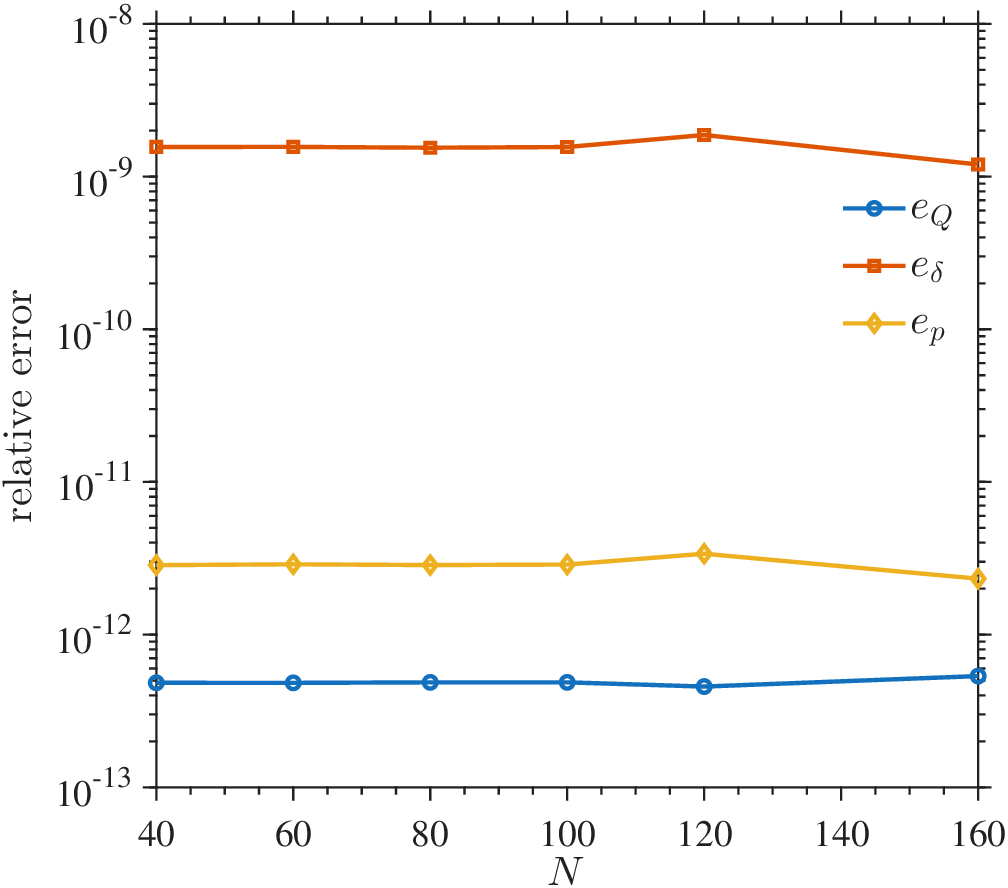}
\caption{}
\label{fig:validation_c}
\end{subfigure}
\caption{Model validation and numerical convergence. (a) Constant-gap benchmark isolating the electrokinetic implementation: computed flux $Q$ matches the exact analytical identity $Q = -s_\zeta$ across all surface conduction values ($Du$). (b) Stiff-wall benchmark isolating the hydrodynamic implementation: the numerically computed augmented-pressure profile $\hat P(x)=P(x)-P(0)$ shows excellent agreement with the exact rigid-channel reference solution from Eq.~\eqref{eq:stiff_ref_Pprime} for varying constriction amplitudes $C$. (c) Grid refinement confirming spectral accuracy: relative errors in flux ($e_Q$), deformation ($e_\delta$), and pressure ($e_p$) remain very small across the tested number of Chebyshev modes $N$.}
\label{fig:validation}
\end{figure}

We validate the solver using exact analytical and asymptotic benchmarks to independently verify the electrokinetic, hydrodynamic, and numerical implementations.First, to verify the correct implementation of the electroosmotic driving and surface conduction terms without the complicating factor of variable geometry, we consider a constant-gap benchmark. By setting $C=0$ and $\delta_s\equiv 0$ so that $h(x)=1$, and removing disjoining forces ($\Pi_{\mathrm{el}}=\Pi_{\mathrm{vdW}}=0$), the augmented pressure gradient vanishes ($dP/dx=0$). The voltage constraint in Eq.~\eqref{eq:nd_field} analytically reduces to $\alpha=1+2Du$, and the flux law in Eq.~\eqref{eq:nd_flux} yields the exact universal identity
\begin{equation}
Q=-s_\zeta
\label{eq:Q_constgap}
\end{equation}
for any $Du$. Figure~\ref{fig:validation}(a) demonstrates agreement within numerical precision between the numerical solver and this exact identity, confirming the global voltage constraint implementation. To isolate and validate the hydrodynamic pressure-flux coupling within a variable geometry, we consider a stiff-wall benchmark. By taking $\mathcal{B}\to\infty$, the wall deformation becomes negligible ($\delta_s\approx 0$) and the gap is strictly defined by the quadratic profile $h\approx h_0=1+C(x-x_c)^2$. In this rigid-channel limit Eq.~\eqref{eq:nd_flux} reduces to
\begin{equation}
Q\approx -\frac{h_0^3}{12}\frac{dP_0}{dx}-s_\zeta\frac{\alpha_0h_0(x)}{h_0(x)+2Du},
\qquad
\alpha_0=\left(\int_0^1\frac{dx}{h_0(x)+2Du}\right)^{-1}.
\label{eq:stiff_ref_flux}
\end{equation}
The reference augmented pressure satisfies
\begin{equation}
P_0'(x)=-\frac{12}{h_0^3}\left(Q_0+s_\zeta\frac{\alpha_0h_0}{h_0+2Du}\right),
\label{eq:stiff_ref_Pprime}
\end{equation}
and it is compared to the numerical result using $\hat P(x)=P(x)-P(0)$. Figure~\ref{fig:validation}(b) compares the numerically integrated augmented pressure $\hat P(x)$ against this analytical reference for several constriction ratios, showing excellent agreement and verifying the spatial discretization of the lubrication terms.  Finally, the mass conservation is verified locally by evaluating the local flux from Eq.~\eqref{eq:nd_flux}
\begin{equation}
q(x)=-\frac{h^3(x)}{12}\frac{dP}{dx}-s_\zeta\frac{\alpha h(x)}{h(x)+2Du},
\label{eq:q_local}
\end{equation}
which should be spatially uniform. We report
\begin{equation}
e_q=\max_{x\in[0,1]}\left|\frac{q(x)}{Q}-1\right|.
\label{eq:eq_error}
\end{equation}
For $Du=0$ and equal end gaps, the identity $Q=-s_\zeta\alpha$ is also recovered numerically. To assess the spectral accuracy of the numerical scheme, grid refinement errors are computed relative to a high-resolution reference solution at polynomial order $N_{\mathrm{ref}}$. The reference fields are interpolated onto the coarse grid using barycentric interpolation on Chebyshev nodes. We report
\begin{equation}
e_Q(N)=\frac{|Q_N-Q_{\mathrm{ref}}|}{|Q_{\mathrm{ref}}|},
\label{eq:eQ_def}
\end{equation}
and weighted $L_2$ profile errors
\begin{equation}
e_\delta(N)=
\left(\frac{\int_0^1\left(\delta_N-\delta_{\mathrm{ref}}\right)^2dx}{\int_0^1\delta_{\mathrm{ref}}^2dx}\right)^{1/2},
\qquad
e_p(N)=
\left(\frac{\int_0^1\left(p_N-p_{\mathrm{ref}}\right)^2dx}{\int_0^1p_{\mathrm{ref}}^2dx}\right)^{1/2}.
\label{eq:profile_errors}
\end{equation}
The integrals in Eq.~\eqref{eq:profile_errors} are evaluated using Clenshaw--Curtis quadrature on the coarse grid. Figure~\ref{fig:validation}(c) and Table 1 demonstrate that the relative errors for flux, deformation, and pressure remain very small across the tested grids, confirming that the Chebyshev collocation method is highly accurate and fully resolved for the chosen range of $N$.

\begin{table}[t]
\centering
\caption{Grid refinement errors computed relative to a reference solution at $N_{\mathrm{ref}}=240$. The polynomial order is $N$ and the number of nodes is $N+1$.}
\label{tab:gridconvergence}
\begin{tabular}{cccc}
\toprule
$N$ & $e_Q$ & $e_\delta$ & $e_p$ \\
\midrule
40  & $4.835\times 10^{-13}$ & $1.564\times 10^{-9}$ & $2.851\times 10^{-12}$ \\
60  & $4.831\times 10^{-13}$ & $1.565\times 10^{-9}$ & $2.887\times 10^{-12}$ \\
80  & $4.867\times 10^{-13}$ & $1.546\times 10^{-9}$ & $2.855\times 10^{-12}$ \\
100 & $4.869\times 10^{-13}$ & $1.562\times 10^{-9}$ & $2.875\times 10^{-12}$ \\
120 & $4.568\times 10^{-13}$ & $1.873\times 10^{-9}$ & $3.387\times 10^{-12}$ \\
160 & $5.351\times 10^{-13}$ & $1.200\times 10^{-9}$ & $2.325\times 10^{-12}$ \\
\bottomrule
\end{tabular}
\end{table}

\section{Results and Discussion}
\label{sec:results}

We present numerical solutions of the fully coupled electroosmotic lubrication and wall bending problem formulated in Sec.~\ref{sec:theory}. Unless otherwise stated, the numerical computations presented in this section utilize the following fixed baseline parameters:constriction apex $x_c=0.5$, scaled inverse Debye length $\kappa^*=30$, repulsive disjoining strength $\Pi_{el}=0.10$, attractive disjoining strength $\Pi_{vdW}=0.01$, baseline dimensionless bending stiffness $\mathcal{B}=0.1$, and identically negatively charged upper and lower walls ($s_\zeta=-1$). In cases where surface conduction is not actively varied, the Dukhin number is fixed at $Du=0.01$. The computations quantify how geometric constriction, wall bending stiffness, surface conduction, and disjoining stresses together control the volumetric flux $Q$ and the minimum gap $h_{\min}=\min_x h(x)$. The observed trends are consistent with the limiting scalings derived in Sec.~\ref{sec:asymptotics}. Over the parameter ranges explored, three qualitative regimes are identified. In the stiff-wall regime the deformation is negligible and $Q$ approaches the rigid-wall value. In the compliance-limited regime the wall deflects near the throat, the gap decreases, and viscous resistance reduces the throughput. In the small-gap saturation regime the gap becomes sufficiently small that the coupled response of bending, hydrodynamic loading, and disjoining pressure causes further thinning to proceed more slowly, leading to a weakly varying finite-flux plateau for very soft walls.

\subsection{Compliance-controlled regimes}
\label{subsec:results_compliance}

\begin{figure}[t]
\centering
\begin{subfigure}{0.32\textwidth}
\centering
\includegraphics[height=5cm,width=\linewidth]{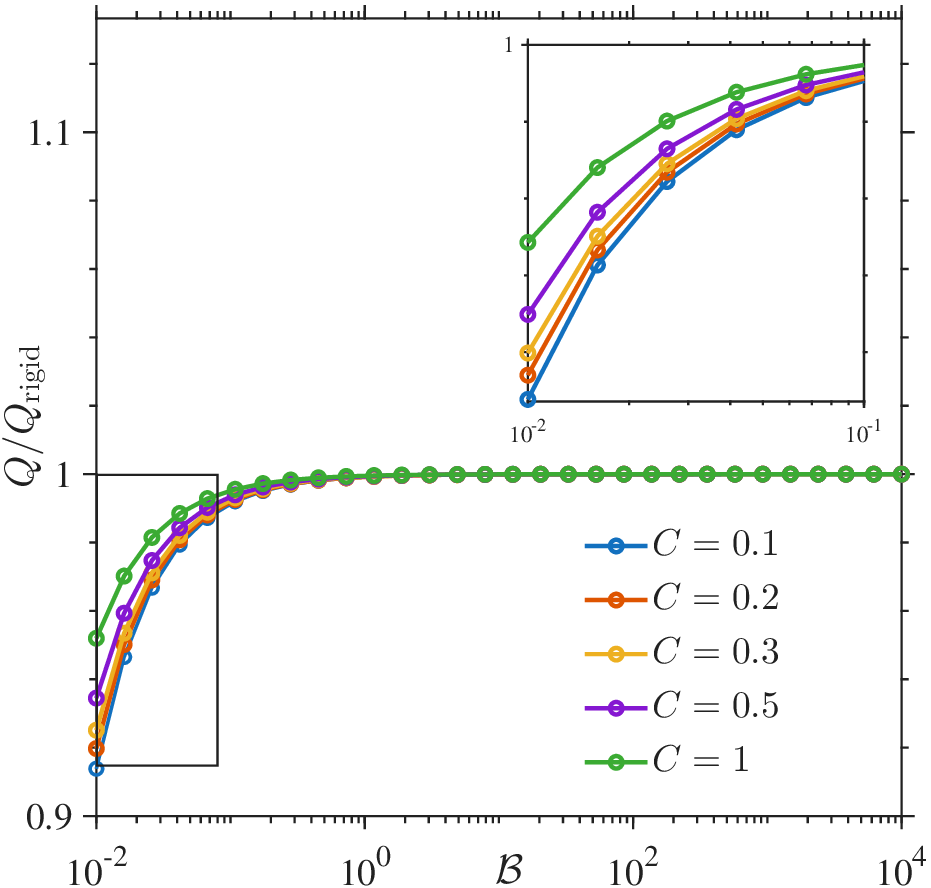}
\caption{}
\label{fig:flux_vs_B_a}
\end{subfigure}
\hfill
\begin{subfigure}{0.32\textwidth}
\centering
\includegraphics[height=5cm,width=\linewidth]{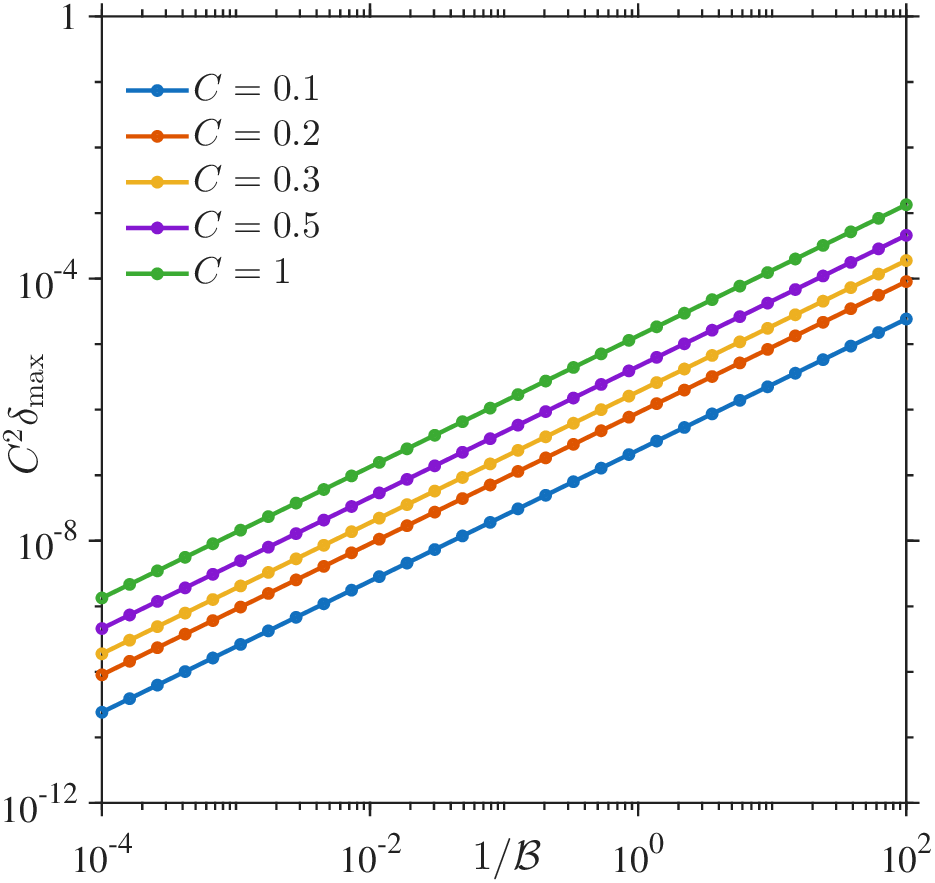}
\caption{}
\label{fig:flux_vs_B_b}
\end{subfigure}
\hfill
\begin{subfigure}{0.32\textwidth}
\centering
\includegraphics[height=5cm,width=\linewidth]{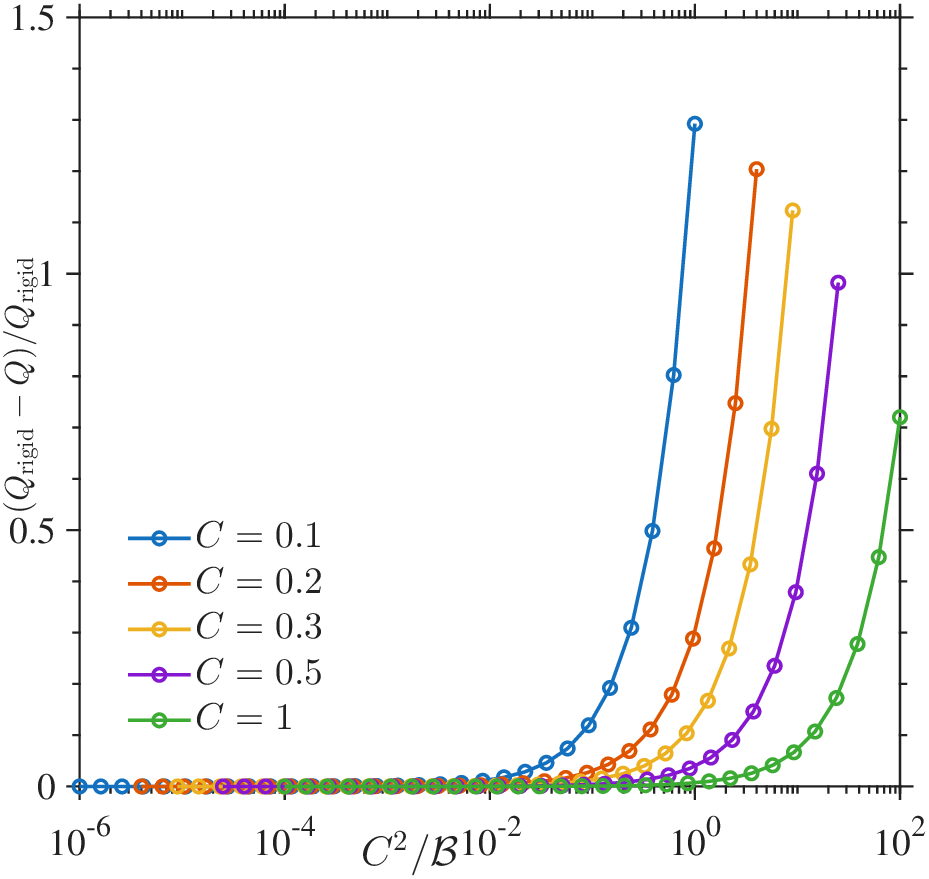}
\caption{}
\label{fig:flux_vs_B_c}
\end{subfigure}
\caption{Compliance-controlled response for several constriction amplitudes $C$. (a) Normalized flux $Q/Q_{\mathrm{rigid}}$ as a function of the dimensionless bending stiffness $\mathcal{B}$. Magnified view of the compliance-limited flux reduction. (b) Maximum deformation magnitude $\delta_{\max}=\max_x |\delta_s(x)|$, shown in the rescaled form $C^2\delta_{\max}$ versus $1/\mathcal{B}$. (c) Normalized flux deficit $(Q_{\mathrm{rigid}}-Q)/Q_{\mathrm{rigid}}$ versus the combined compliance parameter $C^2/\mathcal{B}$.}
\label{fig:flux_vs_B}
\end{figure}

Figure~\ref{fig:flux_vs_B}(a) shows the normalized flux $Q/Q_{\mathrm{rigid}}$ as a function of the bending stiffness $\mathcal{B}$ for several constriction amplitudes $C$. Here $Q_{\mathrm{rigid}}$ denotes the flux computed with the same parameters after setting $\delta_s\equiv 0$. As $\mathcal{B}$ increases, the wall becomes effectively rigid, the deformation weakens, and the flux approaches the rigid-wall limit. As the wall becomes softer, the deformation becomes more pronounced and the throughput decreases relative to the rigid case. For the lower compliant wall considered here, positive hydrodynamic pressure acts downward and therefore tends to increase the local gap. Thus, the observed compliance-induced narrowing of the throat is not caused by hydrodynamic pressure alone; rather, it occurs when the net augmented load $P=p+\Pi(h)$ becomes negative in the constricted region, leading to upward wall motion ($\delta_s>0$) and a reduced gap. This narrowing increases the local hydraulic resistance and thereby reduces the net electroosmotic throughput. Figure~\ref{fig:flux_vs_B}(a) also shows that the constriction amplitude $C$ modulates the onset of this reduction: a sharper constriction localizes the load more strongly, and because the corresponding bending response scales with the factor $C^{-2}$ in Eq.~\eqref{eq:delta_strongC}, the wall behaves as if it were effectively stiffer. Consequently, the compliance-induced flux reduction is shifted to smaller values of $\mathcal{B}$ as $C$ increases.

For sufficiently small $\mathcal{B}$, the flux approaches a weakly varying lower plateau in the numerical results. Within the present model, this saturation reflects the nonlinear coupled balance among elasticity, hydrodynamic loading, and disjoining pressure as the minimum gap becomes small. To avoid over-interpretation, this plateau should be viewed as a numerical saturation of the fully coupled solution for the parameter set considered here, rather than as stand-alone evidence of a universally repulsion-dominated stabilization mechanism. Figure~\ref{fig:flux_vs_B}(b) reports the maximum deformation magnitude in rescaled form, where $\delta_{\max}=\max_x |\delta_s(x)|$. The approximate collapse of $C^2\delta_{\max}$ versus $1/\mathcal{B}$ is consistent with the localized throat balance developed in Sec.~\ref{sec:strongC} and Sec.~\ref{sec:thin_gap_scaling}. In particular, for fixed material and DLVO parameters, the deformation magnitude decreases with increasing $\mathcal{B}$ and with increasing $C^2$, because stronger curvature localizes the load and increases the effective bending resistance through the factor $C^2$ appearing in Eq.~\eqref{eq:delta_strongC}. Figure~\ref{fig:flux_vs_B}(c) shows a corresponding near-collapse of the normalized flux deficit when plotted against $C^2/\mathcal{B}$. This supports the conclusion that, for fixed electrokinetic and intermolecular parameters, the leading compliance effect is governed primarily by the effective stiffness scale $\mathcal{B}C^2$, or equivalently by its inverse $C^2/\mathcal{B}$ used on the horizontal axis. To synthesize these findings, Table \ref{tab:regime_summary} summarizes the primary physical balances, wall deformation characteristics, and throughput behaviors that define each of these three emergent transport regimes.

\begin{table}[htbp]
\centering
\caption{Summary of emergent transport regimes, organized primarily by the effective compliance parameter $C^2/\mathcal{B}$ (equivalently, the effective stiffness $\mathcal{B}C^2$) for fixed electrokinetic and intermolecular parameters.}
\label{tab:regime_summary}
\renewcommand{\arraystretch}{1.5}
\begin{tabular}{p{0.25\linewidth} p{0.3\linewidth} p{0.2\linewidth} p{0.2\linewidth}}
\hline
\textbf{Operating Regime} & \textbf{Primary Physical Balance} & \textbf{Wall Deformation} & \textbf{Throughput Characteristic} \\
\hline
Stiff-wall / effective-stiffness-dominated \newline (small $C^2/\mathcal{B}$; large $\mathcal{B}C^2$) & Wall bending resistance strongly dominates hydrodynamic and disjoining loads. & Negligible \newline ($\delta_s \approx 0$). & Approaches the rigid-wall limit; globally controlled by conductance factor $\alpha$. \\

Compliance-limited \newline (intermediate $C^2/\mathcal{B}$) & Net augmented load becomes negative, driving upward elastic deflection. & Localized narrowing at the throat. & Suppressed significantly due to heightened localized viscous resistance. \\

Small-gap saturation \newline (large $C^2/\mathcal{B}$; small $\mathcal{B}C^2$, within the present parameter range) & Highly nonlinear feedback among bending, hydrodynamic loading, and disjoining pressure as the minimum gap becomes small. & Continues thinning, but at a heavily reduced rate. & Approaches a weakly varying, finite-flux plateau. \\
\hline
\end{tabular}
\end{table}

\subsection{Effect of geometric modulation}
\label{subsec:results_geometry}

\begin{figure}[t]
\centering
\begin{subfigure}{0.32\textwidth}
\centering
\includegraphics[height=5cm,width=\linewidth]{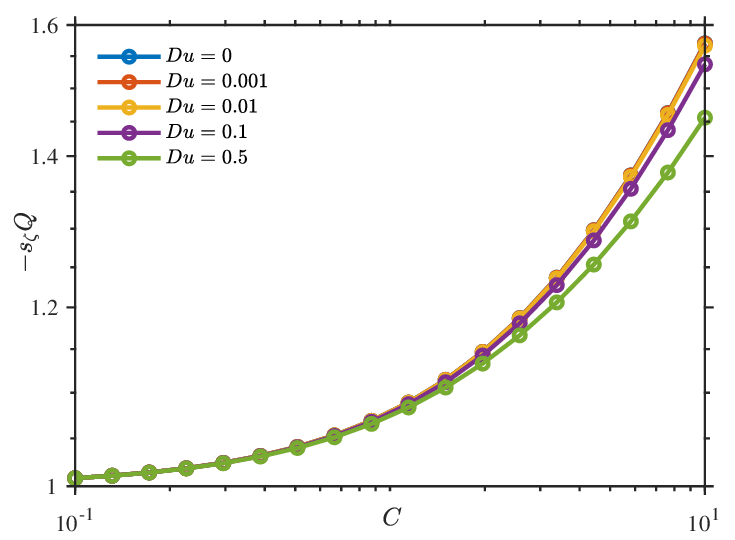}
\caption{}
\label{fig:geometric_focusing_a}
\end{subfigure}
\hfill
\begin{subfigure}{0.32\textwidth}
\centering
\includegraphics[height=5cm,width=\linewidth]{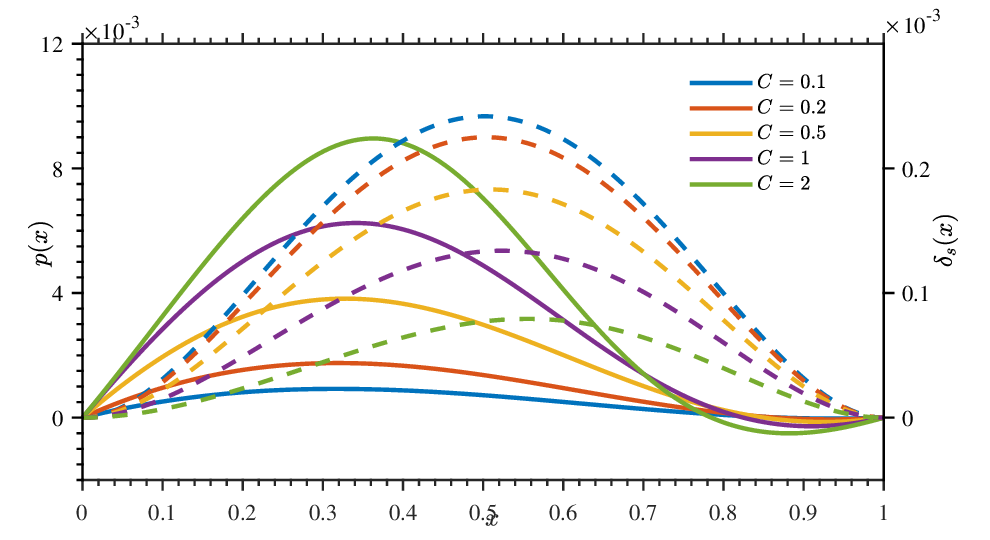}
\caption{}
\label{fig:geometric_focusing_b}
\end{subfigure}
\hfill
\begin{subfigure}{0.32\textwidth}
\centering
\includegraphics[height=5cm,width=\linewidth]{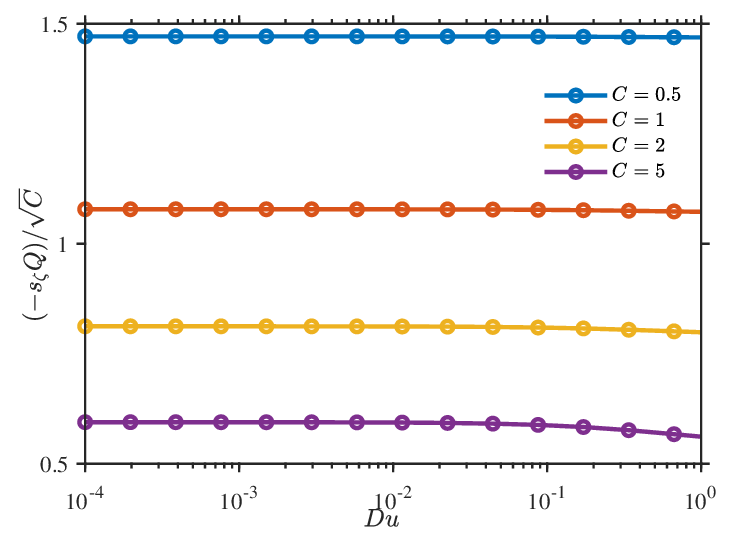}
\caption{}
\label{fig:geometric_focusing_c}
\end{subfigure}

\caption{Effect of geometric modulation and surface conduction. (a) Flux magnitude $-s_\zeta Q$ versus constriction amplitude $C$ on logarithmic axes for several Dukhin numbers $Du$. (b) Representative axial profiles of the hydrodynamic pressure $p(x)$ (solid lines, left axis) and wall deflection $\delta_s(x)$ (dashed lines, right axis) for increasing $C$ at fixed $Du=0.1$. (c) Rescaled flux $(-s_\zeta Q)/\sqrt{C}$ versus $Du$ for different $C$, illustrating the expected trend predicted by the strong-constriction scaling.}
\label{fig:geometric_focusing}
\end{figure}

Geometric curvature $C$ affects the coupled problem through two distinct mechanisms. In the rigid-channel transport problem, increasing $C$ focuses the electric field near the throat and tends to increase the electroosmotic throughput. In the compliant problem, the same increase in $C$ also localizes the load and thereby reduces the deformation sensitivity through the effective stiffness scale $\mathcal{B}C^{2}$. The net compliant throughput therefore reflects a competition between electric-field focusing and compliance-induced hydraulic resistance. The voltage constraint in Eq.~\eqref{eq:nd_field} weights small gaps most strongly, so the electric field becomes larger near the throat as $C$ increases, which strengthens the electroosmotic slip in the constricted region. At the same time, larger $C$ localizes pressure gradients and increases hydrodynamic resistance. The net effect depends on the relative importance of these two mechanisms.

Physically, because the total electric current must be globally conserved, forcing the current through a narrower throat (higher $C$) induces an intense, localized amplification of the electric field. This strong local field acts as a powerful electroosmotic pump at the constriction, successfully overcoming the heightened viscous resistance. Consequently, Figure~\ref{fig:geometric_focusing}(a) shows that the sign-safe flux magnitude $-s_\zeta Q$ increases sublinearly with curvature, following an approximate $\sqrt{C}$ trend over the throat-dominated range. This agrees with the strong constriction scaling in Eq.~\eqref{eq:Q_strongC}. The function $K(Du)$ in Eq.~\eqref{eq:K_closed} captures the reduction of effective electroosmotic mobility due to surface conduction. Figure~\ref{fig:geometric_focusing}(c) shows that plotting $(-s_\zeta Q)/\sqrt{C}$ against $Du$ demonstrates the expected trend consistent with the strong-constriction scaling $K(Du)$. Figure~\ref{fig:geometric_focusing}(b) illustrates the corresponding localization of $p(x)$ and $\delta_s(x)$ near the throat as $C$ increases. The pressure gradient changes sign across the constriction, and the local Poiseuille contribution opposes the electroosmotic plug contribution while maintaining a constant net flux.

\subsection{Narrow gap scalings}
\label{subsec:results_narrowgap}

\begin{figure}[t]
\centering
\begin{subfigure}{0.48\textwidth}
\centering
\includegraphics[width=\linewidth]{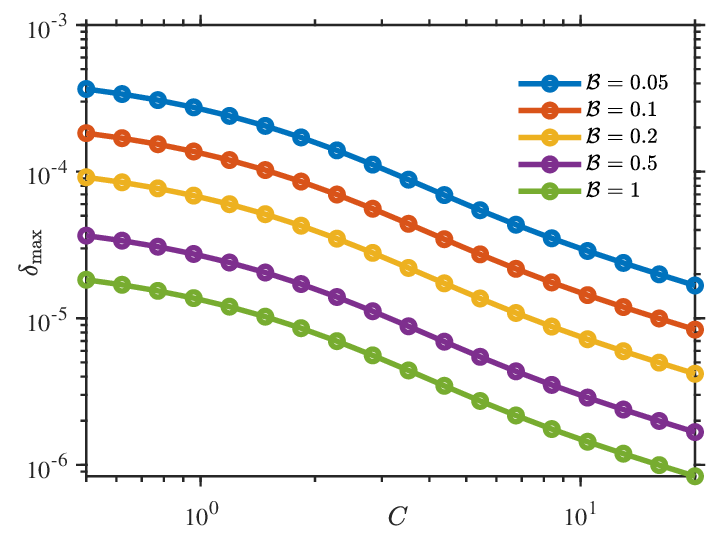}
\caption{}
\label{fig:scaling_results_a}
\end{subfigure}
\hfill
\begin{subfigure}{0.48\textwidth}
\centering
\includegraphics[width=\linewidth]{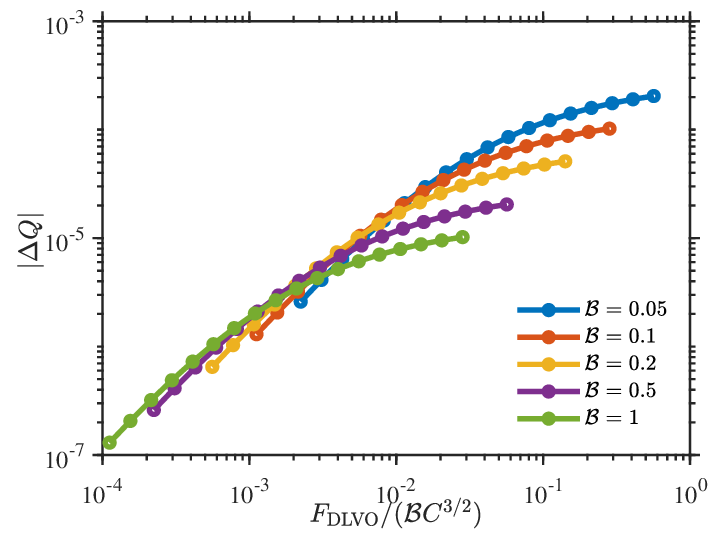}
\caption{}
\label{fig:scaling_results_b}
\end{subfigure}
\caption{Narrow-gap scaling in the DLVO-influenced regime. (a) Maximum deflection $\delta_{\max}$ versus constriction amplitude $C$ for several bending stiffness values $B$, showing the $C^{-2}$ trend. (b) Flux variation $|\Delta Q|$ versus the rescaled forcing $F_{\mathrm{DLVO}}/(B C^{3/2})$, illustrating collapse of the numerical results.}
\label{fig:scaling_results}
\end{figure}

In the narrow gap regime the deformation is controlled by a localized balance between normal traction and bending resistance, while the throughput reduction is controlled by the induced change in electrical conductance. The asymptotic analysis in Sec.~\ref{sec:thin_gap_scaling} predicts the scaling relations in Eq.~\eqref{eq:deltamax_scaling} and Eq.~\eqref{eq:dQ_scaling}. Using the forcing scale $F_{\mathrm{DLVO}}$ defined in Sec.~\ref{sec:thin_gap_scaling}, the deflection magnitude obeys $|\delta|_{\max}\sim F_{\mathrm{DLVO}}/(\mathcal{B}C^2)$ and the throughput variation, defined as $\Delta Q = Q_{rigid} - Q$, obeys $|\Delta Q|\sim F_{\mathrm{DLVO}}/(\mathcal{B}C^{3/2})$.

Figure~\ref{fig:scaling_results}(a) shows that $\delta_{\max}$ decreases with increasing curvature, consistent with the $C^{-2}$ dependence in the localized throat limit. Figure~\ref{fig:scaling_results}(b) shows that the throughput variation collapses when plotted against $F_{\mathrm{DLVO}}/(\mathcal{B}C^{3/2})$. This supports the interpretation that deformation is set by a local balance, while the flux change is controlled by the global conductance correction through $\alpha$.

\subsection{Forcing-dependent response surfaces}
\label{subsec:results_voltage}

In this subsection we vary the applied potential drop through $\Delta\phi=V\,\Delta\phi_0$, where $\Delta\phi_0$ is a fixed reference voltage and $V$ is the dimensionless forcing amplitude. Unlike Sec.~\ref{subsec:nondim}, where the imposed voltage itself sets the velocity and pressure scales, here all quantities are nondimensionalized using the fixed reference scales $U_0=\varepsilon |\zeta|\Delta\phi_0/(\mu L)$ and $p_{\mathrm{ref}}=\mu U_0L/H_0^2$. This choice isolates the effect of forcing, since $C$, $\kappa^\ast$, $Du$, $\Pi_{\mathrm{el}}$, $\Pi_{\mathrm{vdW}}$, and $\mathcal{B}$ remain fixed as $V$ varies. To distinguish this subsection from the baseline scaling of Sec.~\ref{subsec:nondim}, we denote the electric field here by $\mathcal{E}_V(x)=E_xL/\Delta\phi_0$, so that $\int_0^1 \mathcal{E}_V(x)\,dx=V$. The additional notation specific to this subsection is summarized in Appendix A.3. The quasi-one-dimensional current law and the electroosmotic slip scale as,
\begin{equation}
\mathcal{E}_V(x)=\frac{V\,\alpha}{h(x)+2Du},
\qquad
\frac{u_{\mathrm{eo}}}{U_0}=-s_\zeta \mathcal{E}_V(x),
\label{eq:nd_field_V_results}
\end{equation}
and the corresponding dimensionless flux law becomes
\begin{equation}
Q=-\frac{h^3}{12}\frac{dP}{dx}-s_\zeta \frac{V\,\alpha\,h}{h+2Du}.
\label{eq:nd_flux_V_results}
\end{equation}
Thus increasing $V$ increases the electroosmotic forcing at fixed geometry, wall stiffness, and DLVO parameters.

\begin{figure}[!htb]
\centering
\begin{subfigure}{0.48\linewidth}
\includegraphics[height=5.5cm,width=\linewidth]{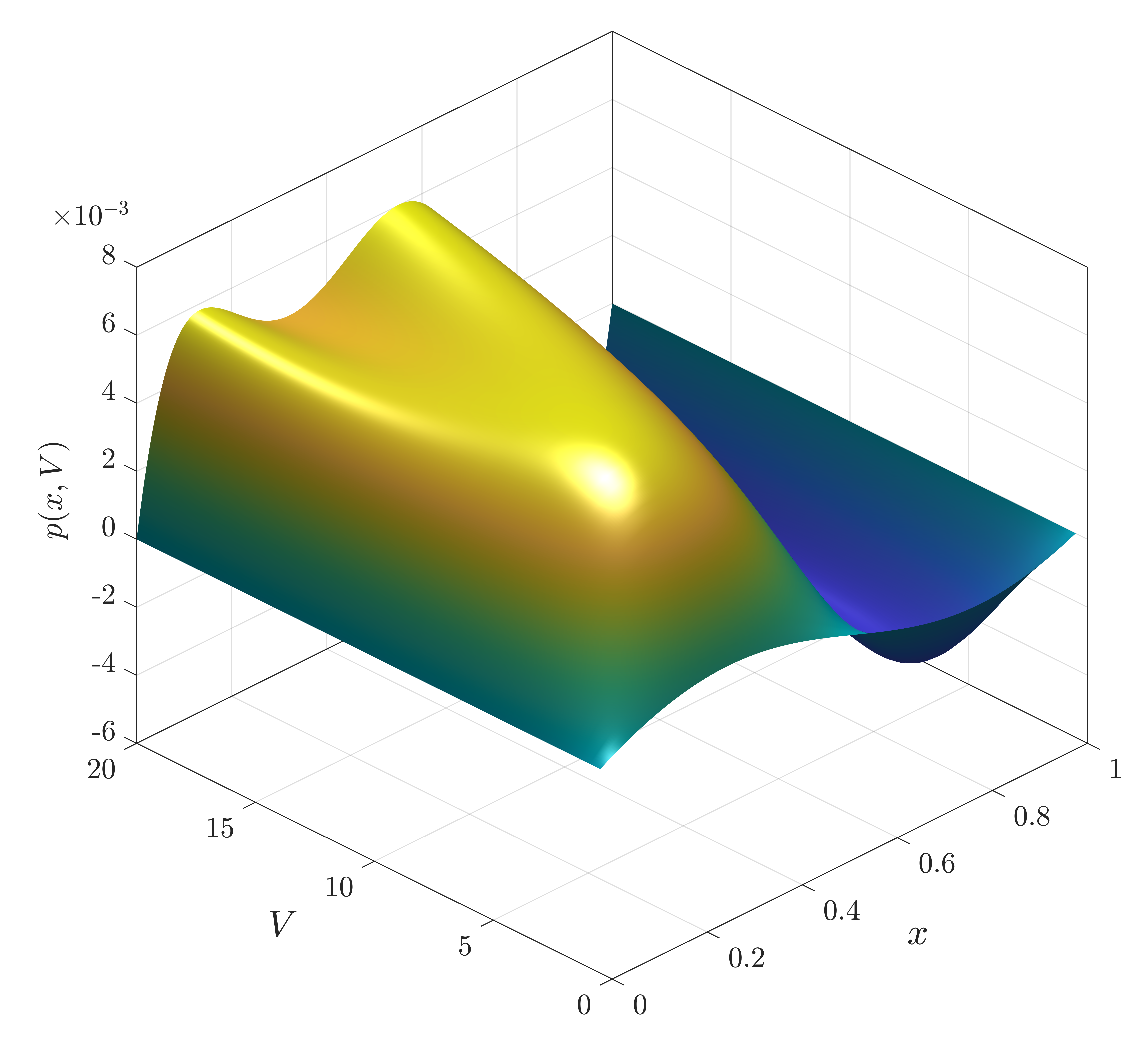}
\end{subfigure}
\hfill
\begin{subfigure}{0.48\linewidth}
\includegraphics[height=5.5cm,width=\linewidth]{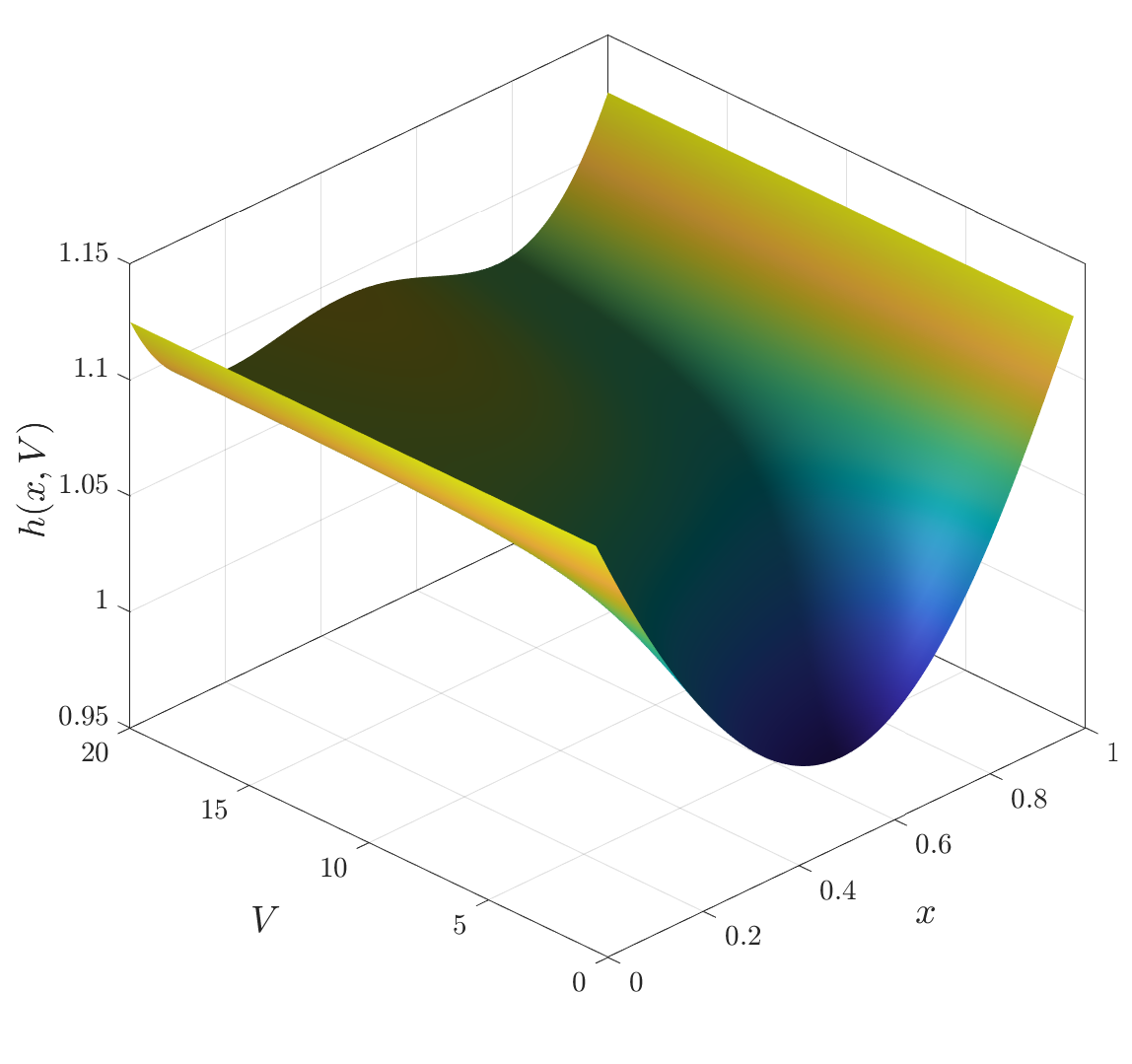}
\end{subfigure}
\caption{Pressure and gap response as functions of axial position and forcing amplitude. In this subsection the forcing amplitude is varied through $V=\Delta\phi/\Delta\phi_0$, where $\Delta\phi_0$ is the fixed reference potential drop used for nondimensionalization. Panel (a) shows the hydrodynamic pressure $p(x,V)$. Panel (b) shows the gap $h(x,V)$. Increasing $V$ strengthens the electroosmotic driving term in Eq.~\eqref{eq:nd_flux_V_results} while the dimensionless groups defined using $\Delta\phi_0$ remain fixed. Other parameters are fixed at the baseline values $x_c=0.5$, $\kappa^\ast=30$, $\Pi_{\mathrm{el}}=0.10$, $\Pi_{\mathrm{vdW}}=0.01$, $\mathcal{B}=0.1$, $Du=0.01$, and $s_\zeta=-1$.}
\label{fig:surfaces}
\end{figure}

\begin{figure}[!ht]
\centering
\includegraphics[width=0.5\linewidth]{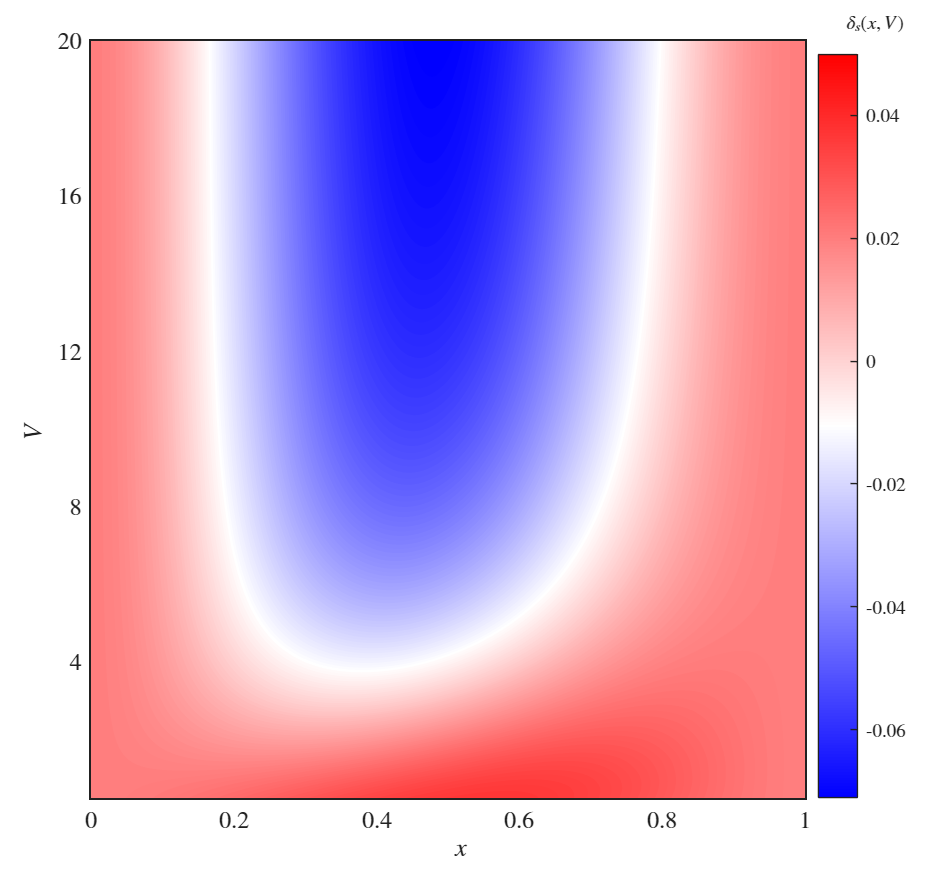}
\caption{Contour map in the $(x,V)$ plane. Color shows the wall displacement $\delta_s(x,V)$ and the overlaid contour lines show constant gap levels $h(x,V)$. The forcing amplitude $V=\Delta\phi/\Delta\phi_0$ controls the strength of the electroosmotic term in Eq.~\eqref{eq:nd_flux_V_results}. At small $V$ the deformation is weak. At intermediate $V$ the response becomes increasingly localized near the throat. At larger $V$ the minimum gap decreases more slowly as the fully coupled balance among bending, hydrodynamic loading, and disjoining pressure becomes increasingly nonlinear. Same parameter set and nondimensionalization as in Figure~\ref{fig:surfaces}.}
\label{fig:surface_plot}
\end{figure}

Figures~\ref{fig:surfaces} and~\ref{fig:surface_plot} summarize the coupled response as the forcing amplitude $V=\Delta\phi/\Delta\phi_0$ is increased at fixed geometry and fixed dimensionless material parameters. Because the present subsection uses the reference scales $U_0$ and $p_{\mathrm{ref}}$, varying $V$ changes only the electroosmotic forcing amplitude and does not redefine $\mathcal{B}$, $\Pi_{\mathrm{el}}$, or $\Pi_{\mathrm{vdW}}$. As $V$ increases, the electroosmotic contribution to the flux law strengthens, and the coupled solution develops larger hydrodynamic pressure gradients, larger wall displacements, and a smaller minimum gap. Figure~\ref{fig:surfaces} shows that the response intensifies primarily in the constricted region near the throat, where geometric focusing amplifies the electric field through Eq.~\eqref{eq:nd_field_V_results}. The stronger local field increases the electroosmotic slip and thereby enhances the pressure gradients required to satisfy the global flux balance. Through the plate equation, these larger loads generate stronger wall deformation and further modify the local gap. The resulting response is therefore highly nonlinear, since the electric field, pressure, wall shape, and hydraulic resistance all depend on the same evolving geometry. Figure~\ref{fig:surface_plot} highlights three qualitative stages. At small $V$ the deformation is weak and the gap varies smoothly with forcing. At intermediate $V$ the pressure and deformation become increasingly localized near the throat and the gap reduction accelerates. At larger $V$ the minimum gap continues to decrease, but at a reduced rate, reflecting the increasingly strong nonlinear feedback among bending, hydrodynamic loading, surface conduction, and disjoining pressure in the narrowest part of the channel.

Together, these results provide a consistent picture of electroosmotic lubrication in compliant constrictions under variable forcing. Geometric focusing increases the local electric field and strengthens the electroosmotic driving. Wall compliance modifies the throat geometry and thereby changes both the hydraulic resistance and the electric-field distribution. Disjoining pressure becomes increasingly important as the gap decreases and contributes to the nonlinear regulation of the minimum gap within the present model. The parameter groups $\mathcal{B}C^2$ and $Du$ therefore remain useful controls for device design through stiffness, curvature, and surface conduction, while $V$ provides an independent measure of the applied electrokinetic forcing.

\subsection{Physical Implications and Limitations}
\label{subsec:limitations}

The coupled mechanics established in this study have direct implications for the design of real-world microfluidic and nanofluidic devices. By deliberately varying the substrate stiffness ($\mathcal{B}$) and geometric curvature ($C$), engineers can design soft-polymeric channels (such as those fabricated from PDMS or hydrogels) that act as self-regulating electroosmotic valves. In such iontronic devices, the applied voltage not only drives the fluid but simultaneously dictates the geometric opening of the valve via elastohydrodynamic feedback, enabling tunable flow control without moving mechanical components. However, translating this theoretical framework to real-world applications requires navigating several physical limitations. First, our model strictly assumes thin, nonoverlapping EDLs and quasi-Ohmic conduction. If the soft wall deforms so that the gap approaches the nanoscale (comparable to the Debye length), the thin-EDL approximation breaks down. In such extremely narrow confinements, the assumption of a constant zeta potential is no longer strictly valid; instead, models must incorporate dynamic charge regulation, wherein the surface ionization state chemically adjusts to local ion concentrations and overlapping potentials.

Furthermore, the experimental realization faces distinct thermomechanical challenges. The intense electric field focusing at the throat ($C \gg 1$), which is responsible for the enhanced throughput, inevitably generates highly localized Joule heating. This thermal gradient can alter fluid viscosity, modify the local substrate stiffness, and induce the unwanted thermoviscous flows. Finally, while we model the wall as an ideal elastic Kirchhoff--Love plate, real soft polymers exhibit viscoelastic hysteresis and nonlinear strain stiffening, which could introduce dynamic time delays in the valve response under transient voltage operations.

\section{Conclusions}
\label{sec:conclusion}

We developed a nonlinear model for electroosmotic lubrication in a constricted microchannel with a rigid curved upper wall and a compliant lower wall. The formulation couples Helmholtz--Smoluchowski slip, lubrication hydrodynamics, Kirchhoff--Love plate bending, and disjoining pressure, with electrical resistance determined by quasi-one-dimensional conduction including surface conductance. Numerical solutions supported by asymptotic analysis reveal three operating regimes. In the stiff-wall regime, deformation is negligible and the throughput approaches the rigid-wall value controlled by the conductance factor $\alpha$ and the Dukhin number. In the compliance-limited regime, bending-induced throat narrowing increases hydrodynamic resistance and reduces throughput. In the small-gap saturation regime, the minimum gap becomes sufficiently small that the coupled response of bending, hydrodynamic loading, and disjoining pressure causes further thinning to proceed more slowly, yielding a weakly varying finite-flux plateau for very soft walls.

In the strong-constriction limit, compact scaling relations organize the coupled response. The conductance factor scales as $\alpha\sim \sqrt{C}$ and the throughput obeys $Q\sim -s_\zeta\sqrt{C}\,K(Du)$, where $K(Du)$ quantifies the reduction of effective electroosmotic mobility by surface conduction. The wall response is localized near the throat and the maximum deflection scales as $|\delta_s|_{\max}\sim P_{\max}/(\mathcal{B}C^2)$, while the corresponding throughput variation scales as $|\Delta Q|\sim F_{\mathrm{DLVO}}/(\mathcal{B}C^{3/2})$, where $F_{\mathrm{DLVO}}$ is a characteristic disjoining forcing based on the magnitude of $\Pi(h_{\min})$. These relations are consistent with the numerical results over the explored parameter ranges and show that curvature controls electrical focusing and localization, bending stiffness controls deformation amplitude, and disjoining pressure contributes to regulating the minimum gap.

The composite groups $\mathcal{B}C^2$ and $Du$ emerge as the main design-relevant control parameters for soft electroosmotic channels. Increasing $\mathcal{B}$ suppresses deformation and reduces compliance induced flux losses. The increment in $Du$ has a dual effect. It  increases the rigid channel throughput through electric field focusing, but it also localizes the load and thereby reduces the deformation sensitivity through $\mathcal{B}C^2$. Increase in $Du$ reduces the electroosmotic throughput through enhanced surface conduction. Within the thin, nonoverlapping double-layer and quasi-Ohmic conduction assumptions, the present framework provides a consistent basis for interpreting compliant electrokinetic constrictions and for developing design-oriented scaling laws. The compact asymptotic rules derived here provide a quantitative basis for future experimental studies in soft electrokinetics.

	\section*{Acknowledgments}
	
	S. Sahoo gratefully acknowledges the financial support received from the Ministry of Human Resource Development (MHRD), Government of India. A. K. Nayak thanks the NBHM, India, and the Science and Engineering Research Board (SERB), India (Grant No. CRG/2023/006863), for their support during the preparation of this manuscript.
	
	\section*{Data Availability}
	
	The data that support the findings of this study are available within the article.
	
	\section*{Author Notes}
	The authors declare that they have no competing interests.

	\appendix
	\section*{Appendix}
\renewcommand{\thesection}{A}
\renewcommand{\thesubsection}{A.\arabic{subsection}}
\renewcommand{\theequation}{A.\arabic{equation}}
\setcounter{equation}{0}

\subsection{Matched asymptotics for electroosmotic slip in the thin-EDL limit}

This appendix gives a brief justification of the Helmholtz--Smoluchowski slip condition in Eq.~\eqref{eq:ueo} using the standard thin-EDL argument. We follow classical electrokinetic theory \cite{probstein2005physicochemical,hunter2013zeta}. Consider steady unidirectional flow between parallel plates with a wall at $y=0$ and the bulk at $y\to\infty$. The electrolyte is symmetric $z:z$ with constant viscosity $\mu$ and permittivity $\varepsilon$. Let $\psi(y)$ denote the equilibrium EDL potential, with $\psi(0)=\zeta$ and $\psi(\infty)=0$. The applied axial potential varies slowly in $x$, so the imposed axial potential gradient $G=\frac{d\phi}{dx}$ is effectively uniform across the EDL thickness. In Sec.~\ref{sec:theory} the axial electric field is defined by $E_x=-\frac{d\phi}{dx}$, so that $G=-E_x$.

The inner coordinate is $\eta=y/\lambda_D$, and the dimensionless Debye length is $\delta=\lambda_D/H_0\ll 1$. The equilibrium potential satisfies the Poisson--Boltzmann equation in the inner region, although its explicit form is not needed for the leading-order slip result. The charge density is related to the potential by the Poisson equation
\begin{equation}
\varepsilon \frac{d^2\psi}{dy^2}=-\rho_e(y).
\label{eq:poisson_app}
\end{equation}
Neglecting any imposed axial pressure gradient within the EDL, the axial Stokes balance is
\begin{equation}
0=\mu \frac{d^2u}{dy^2}+\rho_e(y)E_x.
\label{eq:stokes_app}
\end{equation}
Since $E_x=-G$, this can be written equivalently as
\begin{equation}
\mu \frac{d^2u}{dy^2}=\rho_e(y)\,G.
\label{eq:stokes_app_G}
\end{equation}
Combining \eqref{eq:poisson_app} and \eqref{eq:stokes_app_G} gives
\begin{equation}
\mu \frac{d^2u}{dy^2}
=-\varepsilon G \frac{d^2\psi}{dy^2}.
\label{eq:stokes_poisson_app}
\end{equation}
Integrating once and using matching to the outer region, for which $u_y\to 0$ and $\psi_y\to 0$ as $y\to\infty$, yields
\begin{equation}
\mu \frac{du}{dy}
=-\varepsilon G \frac{d\psi}{dy}.
\end{equation}
Integrating again gives
\begin{equation}
\mu u
=-\varepsilon G \psi+\text{const}.
\end{equation}
The no-slip condition at the wall implies $u(0)=0$ and $\psi(0)=\zeta$, so $\text{const}=\varepsilon G\zeta$. Therefore,
\begin{equation}
u(y)=\frac{\varepsilon G}{\mu}\big(\zeta-\psi(y)\big).
\end{equation}
At the outer edge of the EDL, where $\psi(\infty)=0$, the matched slip velocity is
\begin{equation}
u(\infty)=\frac{\varepsilon G\zeta}{\mu}.
\end{equation}
Using $G=-E_x$ recovers the Helmholtz--Smoluchowski slip law in the form used in Sec.~\ref{sec:theory},
\begin{equation}
u_{\mathrm{slip}}=-\frac{\varepsilon \zeta}{\mu}E_x.
\end{equation}
Corrections to this leading-order slip are smaller by $O(\lambda_D/H_0)$ and are negligible when $\kappa^\ast=H_0/\lambda_D\gg 1$.

\subsection{Derivation of the first order flux correction in Eq.~\eqref{eq:stiff_q1}}
\label{app:q1_derivation}

We start from the dimensionless flux law
\[
Q=-\frac{h^3(x)}{12}\frac{dP}{dx}-s_\zeta\frac{\alpha h(x)}{h(x)+2Du},
\qquad
\alpha=\left(\int_0^1\frac{dx}{h(x)+2Du}\right)^{-1}.
\]
In the stiff-wall limit, define $\epsilon=\mathcal{B}^{-1}\ll 1$ and expand
\[
p=p_0+\epsilon p_1+\cdots,\qquad
\delta_s=\epsilon\delta_1+\cdots,\qquad
h=h_0-\epsilon\delta_1+\cdots,\qquad
P=P_0+\epsilon P_1+\cdots,\qquad
\alpha=\alpha_0+\epsilon\alpha_1+\cdots.
\]
The leading order gap is $h_0(x)=1+C(x-x_c)^2$. The augmented pressure is
\[
P=p+\Pi_{\mathrm{el}}e^{-\kappa^\ast h}-\frac{\Pi_{\mathrm{vdW}}}{h^3}.
\]
Expanding the disjoining contribution about $h_0$ gives
\[
\Pi_{\mathrm{el}}e^{-\kappa^\ast h}-\frac{\Pi_{\mathrm{vdW}}}{h^3}
=
\left(\Pi_{\mathrm{el}}e^{-\kappa^\ast h_0}-\frac{\Pi_{\mathrm{vdW}}}{h_0^3}\right)
+\epsilon\delta_1\left(\kappa^\ast\Pi_{\mathrm{el}}e^{-\kappa^\ast h_0}-\frac{3\Pi_{\mathrm{vdW}}}{h_0^4}\right)
+O(\epsilon^2).
\]
Therefore
\[
P_0=p_0+\Pi_{\mathrm{el}}e^{-\kappa^\ast h_0}-\frac{\Pi_{\mathrm{vdW}}}{h_0^3},
\qquad
P_1=p_1+\delta_1K_{\mathrm{DLVO}},
\]
with
\[
K_{\mathrm{DLVO}}(x)=\kappa^\ast\Pi_{\mathrm{el}}e^{-\kappa^\ast h_0(x)}-\frac{3\Pi_{\mathrm{vdW}}}{h_0^4(x)}.
\]

\paragraph{Poiseuille contribution}
Using $(h_0-\epsilon\delta_1)^3=h_0^3-3\epsilon h_0^2\delta_1+O(\epsilon^2)$ and $P'=P_0'+\epsilon P_1'+\cdots$, we obtain
\[
-\frac{h^3}{12}P'
=
-\frac{h_0^3}{12}P_0'
+\epsilon\left(-\frac{h_0^3}{12}P_1'+\frac{h_0^2\delta_1}{4}P_0'\right)
+O(\epsilon^2).
\]

\paragraph{Electroosmotic contribution}
Expand the electroosmotic term about $h_0$ and $\alpha_0$
\[
-s_\zeta\frac{\alpha h}{h+2Du}
=
-s_\zeta\frac{\alpha_0 h_0}{h_0+2Du}
+\epsilon s_\zeta\left(\delta_1\frac{2\alpha_0 Du}{(h_0+2Du)^2}-\frac{h_0}{h_0+2Du}\alpha_1\right)
+O(\epsilon^2),
\]
since $h=h_0-\epsilon\delta_1$.

\paragraph{Voltage constraint and $\alpha_1$}
From
\[
\frac{1}{h+2Du}=\frac{1}{h_0+2Du}+\epsilon\frac{\delta_1}{(h_0+2Du)^2}+O(\epsilon^2),
\]
write $I=\int_0^1(h+2Du)^{-1}dx=I_0+\epsilon I_1+\cdots$ with
\[
I_0=\int_0^1\frac{dx}{h_0+2Du},
\qquad
I_1=\int_0^1\frac{\delta_1(x)}{(h_0(x)+2Du)^2}dx.
\]
Then $\alpha=1/I$ gives
\[
\alpha_0=\frac{1}{I_0},
\qquad
\alpha_1=-\alpha_0^2 I_1
=-\alpha_0^2\int_0^1\frac{\delta_1(x)}{(h_0(x)+2Du)^2}dx.
\]

\paragraph{Collecting orders}
Combining the $O(1)$ and $O(\epsilon)$ terms gives
\[
Q=Q_0+\epsilon q_1(x)+O(\epsilon^2),
\]
with
\[
Q_0=-\frac{h_0^3}{12}P_0'-s_\zeta\frac{\alpha_0 h_0}{h_0+2Du},
\]
and
\[
q_1(x)
=
-\frac{h_0^3}{12}P_1'(x)
+\frac{h_0^2\delta_1}{4}P_0'(x)
+s_\zeta\left(\delta_1(x)\frac{2\alpha_0 Du}{(h_0+2Du)^2}-\frac{h_0}{h_0+2Du}\alpha_1\right),
\]
which is Eq.~\eqref{eq:stiff_q1}. The flux $Q$ is spatially uniform, so the constant correction $Q_1$ is obtained by enforcing a compatibility condition, for example by averaging $q_1$ over $x\in[0,1]$. For equal end gaps the boundary contributions associated with $P_1'$ cancel. In the Ohmic limit $Du\to 0$, the local electroosmotic correction proportional to $Du$ vanishes and the remaining correction enters through $\alpha_1$.

	\subsection{Nomenclature}
    \subsection*{Dimensionless variables and parameters}
        \begin{tabular}{lll}
            \hline
            Symbol & Description & Definition \\
            \hline
            $x$ & Axial coordinate & $x/L$ \\
            $x_c$ & Constriction location & $X_c/L$ \\
            $h$ & Gap & $h/H_0$ \\
            $\delta_s$ & Wall deflection & $\delta_s/H_0$ \\
            $p$ & Hydrodynamic pressure & $p/p_\star$ \\
            $P$ & Augmented pressure & $P/p_\star$ \\
            $q$ & Flux & $q/(U_\star H_0)$ \\
            $Q$ & Constant flux & $Q/(U_\star H_0)$ \\
            $C$ & Curvature parameter & $L^2/(2RH_0)$ \\
            $\kappa^\ast$ & Scaled inverse Debye length & $\kappa H_0$ \\
            $Du$ & Dukhin number & $K_s/(\sigma_b H_0)$ \\
            $\Pi_{\mathrm{el}}$ & Repulsive disjoining strength & $64n_0k_BT\tanh^2\!\left(\frac{ze\psi_0}{4k_BT}\right)/p_\star$ \\
            $\Pi_{\mathrm{vdW}}$ & Attractive disjoining strength & $\left(A_H/(6\pi H_0^3)\right)/p_\star$ \\
            $\mathcal{B}$ & Bending stiffness & $D H_0^3/(\mu U_\star L^5)$ \\
            $s_\zeta$ & Zeta sign factor & $\zeta/|\zeta|$ \\
            $\mathcal{E}(x)$ & Dimensionless electric field & $E_x L/\Delta\phi$ \\
            $\alpha$ & Conductance factor & $\left(\int_0^1\frac{dx}{h(x)+2Du}\right)^{-1}$ \\
            \hline
        \end{tabular}

    \subsection*{Dimensional symbols}
        \begin{tabular}{lll}
            \hline
            Symbol & Description & Units \\
            \hline
            $x$ & Axial coordinate & m \\
            $L$ & Channel length & m \\
            $X_c$ & Location of constriction apex & m \\
            $h(x)$ & Local gap height & m \\
            $H_0$ & Minimum undeformed gap & m \\
            $R$ & Radius of curvature of upper wall & m \\
            $\delta_s(x)$ & Wall deflection measured positive upward into the fluid & m \\
            $p(x)$ & Hydrodynamic pressure & Pa \\
            $P(x)$ & Augmented pressure $p+\Pi(h)$ & Pa \\
            $\Pi(h)$ & Disjoining pressure & Pa \\
            $q(x)$ & Volumetric flux per unit width & m$^2$/s \\
            $Q$ & Constant flux & m$^2$/s \\
            $u_{\mathrm{eo}}(x)$ & Electroosmotic slip velocity & m/s \\
            $U_\star$ & Reference EO velocity scale $\varepsilon|\zeta|\Delta\phi/(\mu L)$ & m/s \\
            $E_x(x)$ & Axial electric field & V/m \\
            $\phi(x)$ & Electric potential & V \\
            $\Delta \phi$ & Applied potential drop & V \\
            $\zeta$ & Zeta potential & V \\
            $\varepsilon$ & Permittivity & F/m \\
            $\mu$ & Dynamic viscosity & Pa s \\
            $\sigma_b$ & Bulk conductivity & S/m \\
            $K_s$ & Surface conductance per wall & S \\
            $z$ & Ion valence & 1 \\
            $n_0$ & Bulk ion concentration & m$^{-3}$ \\
            $k_B$ & Boltzmann constant & J/K \\
            $T$ & Temperature & K \\
            $e$ & Elementary charge & C \\
            $\psi_0$ & Stern-plane potential & V \\
            $\kappa$ & Inverse Debye length & m$^{-1}$ \\
            $\lambda_D$ & Debye length & m \\
            $A_H$ & Hamaker constant & J \\
            $h_s$ & Wall thickness & m \\
            $E_Y$ & Young's modulus & Pa \\
            $\nu$ & Poisson ratio & 1 \\
            $D$ & Plate flexural rigidity & N m \\
            \hline
        \end{tabular}

    \subsection*{Auxiliary and asymptotic notation}
        \begin{tabular}{ll}
            \hline
            Symbol & Meaning \\
            \hline
            Flux law & $Q=-\frac{h^3}{12}\frac{dP}{dx}-s_\zeta\frac{\alpha h}{h+2Du}$ \\
            Plate balance & $P(x)=-\mathcal{B}\frac{d^4\delta_s}{dx^4}$ \\
            $\epsilon$ & Small parameter $1/\mathcal{B}$ \\
            $h_0(x)$ & Base gap $1+C(x-x_c)^2$ \\
            $p_0,p_1$ & Asymptotic hydrodynamic pressures \\
            $P_0,P_1$ & Asymptotic augmented pressures \\
            $\delta_1$ & First order wall deflection in the stiff wall expansion \\
            $K_{\mathrm{DLVO}}$ & Incremental disjoining stiffness $\kappa^\ast\Pi_{\mathrm{el}}e^{-\kappa^\ast h_0}-3\Pi_{\mathrm{vdW}}/h_0^4$ \\
            $\xi$ & Stretched coordinate $\sqrt{C}(x-x_c)$ \\
            $F_{\mathrm{DLVO}}$ & Forcing scale $\max\!\left(\Pi_{\mathrm{el}}e^{-\kappa^\ast h_{\min}},\Pi_{\mathrm{vdW}}/h_{\min}^3\right)$ \\
            $h_{\min}$ & Minimum gap in the throat region \\
            $K(Du)$ & Mobility function in the strong constriction limit \\
            \hline
        \end{tabular}
    
    \subsection*{Additional notation used in Sec. 5.4}
    
        \begin{tabular}{lll}
            \hline
            Symbol & Description & Definition \\
            \hline
            $V$ & Forcing amplitude & $\Delta\phi/\Delta\phi_0$ \\
            $\Delta\phi_0$ & Reference potential drop & \\
            $U_0$ & Reference electroosmotic velocity & $\varepsilon |\zeta|\Delta\phi_0/(\mu L)$ \\
            $p_{\mathrm{ref}}$ & Reference pressure & $\mu U_0L/H_0^2$ \\
            $\mathcal{E}_V(x)$ & Dimensionless electric field in Sec. 5.4 & $E_xL/\Delta\phi_0$ \\
            \hline
        \end{tabular}
    \vspace{0.2cm}
    
    \noindent In Sec.~\ref{subsec:results_voltage}, $p$, $P$, $q$, and $Q$ are nondimensionalized using $p_{\mathrm{ref}}$ and $U_0H_0$, rather than the baseline scales $p_\ast$ and $U_\ast H_0$ used elsewhere.


\begin{thebibliography}{References}
		\small
		
		\bibitem{saville1977electrokinetic} Saville, D. A. (1977). Electrokinetic effects with small particles. Annual Review of Fluid Mechanics, 9(1), 321-337.
		
		\bibitem{ren2001interfacial} Ren, L., Qu, W., \& Li, D. (2001). Interfacial electrokinetic effects on liquid flow in microchannels. International Journal of Heat and Mass Transfer, 44(16), 3125-3134.
		
		\bibitem{probstein2005physicochemical}Probstein, R. F. (2005). Physicochemical hydrodynamics: an introduction. John Wiley \& Sons.
		
		\bibitem{hunter2013zeta}Hunter, R. J. (2013). Zeta potential in colloid science: principles and applications (Vol. 2). Academic Press.
		
		\bibitem{schnitzer2012induced}Schnitzer, O., \& Yariv, E. (2012). Induced-charge electro-osmosis beyond weak fields. Physical Review E—Statistical, Nonlinear, and Soft Matter Physics, 86(6), 061506.
		
		\bibitem{wong2004electrokinetics} Wong, P. K., Wang, T. H., Deval, J. H., \& Ho, C. M. (2004). Electrokinetics in micro devices for biotechnology applications. IEEE/ASME transactions on mechatronics, 9(2), 366-376.
		
		\bibitem{fiechtner2003faceted} Fiechtner, G. J., \& Cummings, E. B. (2003). Faceted design of channels for low-dispersion electrokinetic flows in microfluidic systems. Analytical Chemistry, 75(18), 4747-4755.
		
		\bibitem{fernandez2021concentration}Fern{\'a}ndez-Mateo, R., Calero, V., Morgan, H., Ramos, A., \& Garc{\'\i}a-S{\'a}nchez, P. (2021). Concentration–polarization electroosmosis near insulating constrictions within microfluidic channels. Analytical Chemistry, 93(44), 14667-14674.
		
		\bibitem{koyama2021electro}Koyama, S., Inoue, D., Okada, A., \& Yoshida, H. (2021). Electro-osmotic diode based on colloidal nano-valves between double membranes. Physical Review Research, 3(3), 033289.
		
		\bibitem{gu2022electrokinetics} Gu, Z., Huo, P., Xu, B., Su, M., Bazant, M. Z., \& Deng, D. (2022). Electrokinetics in two-dimensional complicated geometries: Conformal mapping and experimental comparison. Physical Review Fluids, 7(3), 033701.
		
		\bibitem{chakraborty2012fluid} Chakraborty, D., Prakash, J. R., Friend, J., \& Yeo, L. (2012). Fluid-structure interaction in deformable microchannels. Physics of Fluids, 24(10).
		
		\bibitem{yu2013deformable} Yu, H., \& Zhou, G. (2013). Deformable mold based on-demand microchannel fabrication technology. Sensors and Actuators B: Chemical, 183, 40-45.
		
		\bibitem{ozsun2013non} Ozsun, O., Yakhot, V., \& Ekinci, K. L. (2013). Non-invasive measurement of the pressure distribution in a deformable micro-channel. Journal of Fluid Mechanics, 734, R1.
		
		\bibitem{mehboudi2018one} Mehboudi, A., \& Yeom, J. (2018). A one-dimensional model for compressible fluid flows through deformable microchannels. Physics of Fluids, 30(9).
		
		\bibitem{wang2019theory} Wang, X., \& Christov, I. C. (2019). Theory of the flow-induced deformation of shallow compliant microchannels with thick walls. Proceedings of the Royal Society A, 475(2231), 20190513.
		
		\bibitem{guyard2022elastohydrodynamic}Guyard, G., Restagno, F., \& McGraw, J. D. (2022). Elastohydrodynamic relaxation of soft and deformable microchannels. Physical Review Letters, 129(20), 204501.
		
		\bibitem{greidanus2022response}Greidanus, A. J., Delfos, R., Picken, S. J., \& Westerweel, J. (2022). Response regimes in the fluid–structure interaction of wall turbulence over a compliant coating. Journal of Fluid Mechanics, 952, A1.
		
		\bibitem{roy2024fluid} Roy, A., \& Dhar, P. (2024). Fluid–structure-interactive elasto-and thermo-hydrodynamics of electrokinetic binary fluid flows in compliant micro-confinements. Physics of Fluids, 36(3).
		
		\bibitem{essink2021regimes}Essink, M. H., Pandey, A., Karpitschka, S., Venner, C. H., \& Snoeijer, J. H. (2021). Regimes of soft lubrication. Journal of Fluid Mechanics, 915, A49.
		
		\bibitem{karan2021generalization}Karan, P., Chakraborty, J., \& Chakraborty, S. (2021). Generalization of elastohydrodynamic interactions between a rigid sphere and a nearby soft wall. Journal of Fluid Mechanics, 923, A32.
		
		\bibitem{ghosal2002lubrication}Ghosal, S. (2002). Lubrication theory for electro-osmotic flow in a microfluidic channel of slowly varying cross-section and wall charge. Journal of Fluid Mechanics, 459, 103-128.
		
		\bibitem{shit2024stability}Shit, G. C., Sengupta, A., \& Mondal, P. K. (2024). Stability analysis of electro-osmotic flow in a rotating microchannel. Journal of Fluid Mechanics, 983, A13.
		
		\bibitem{goyal2024generalizing}Goyal, V., Datta, S., \& Chakraborty, S. (2024). Generalizing electroosmotic-flow predictions over charge-modulated periodic topographies: tuneable far-field effects. Journal of Fluid Mechanics, 990, A1.
		
		\bibitem{skotheim2005soft} Skotheim, J. M., \& Mahadevan, L. (2005). Soft lubrication: The elastohydrodynamics of nonconforming and conforming contacts. Physics of Fluids, 17(9).
		
		\bibitem{skotheim2004dynamics} Skotheim, J. M., \& Mahadevan, L. (2004). Dynamics of poroelastic filaments. Proceedings of the Royal Society A, 460(2047), 1995-2020.
		
		\bibitem{skotheim2004soft}Skotheim, J. M., \& Mahadevan, L. (2004). Soft lubrication. Physical Review Letters, 92(24), 245509.
		
		\bibitem{salez2015elastohydrodynamics} Salez, T., \& Mahadevan, L. (2015). Elastohydrodynamics of a sliding, spinning and sedimenting cylinder near a soft wall. Journal of Fluid Mechanics, 779, 181-196.
		
		\bibitem{bertin2022soft} Bertin, V., Amarouchene, Y., Rapha{\"e}l, E., \& Salez, T. (2022). Soft-lubrication interactions between a rigid sphere and an elastic wall. Journal of Fluid Mechanics, 933, A23.
		
		\bibitem{rallabandi2024fluid} Rallabandi, B. (2024). Fluid-elastic interactions near contact at low Reynolds number. Annual Review of Fluid Mechanics, 56(1), 491-519.
		
		\bibitem{bharti2024singular} Bharti, Ferreira, Q., Jha, A., Carlson, A., Dean, D. S., Amarouchene, Y., Chan, T. S., \& Salez, T. (2024). Singular viscoelastic perturbation to soft lubrication. Physical Review Research, 6(4), 043060.
		
		\bibitem{paludan2024elastohydrodynamic}Paludan, M. V., Dollet, B., Marmottant, P., \& Jensen, K. H. (2024). Elastohydrodynamic interactions in soft hydraulic knots. Journal of Fluid Mechanics, 984, A55.
		
		\bibitem{fares2024observation}Fares, N., Lavaud, M., Zhang, Z., Jha, A., Amarouchene, Y., \& Salez, T. (2024). Observation of Brownian elastohydrodynamic forces acting on confined soft colloids. Proceedings of the National Academy of Sciences, 121(42), e2411956121.		
		
		\bibitem{wang2015out} Wang, Y., Dhong, C., \& Frechette, J. (2015). Out-of-contact elastohydrodynamic deformation due to lubrication forces. Physical Review Letters, 115(24), 248302.
		
		\bibitem{davies2018elastohydrodynamic} Davies-Strickleton, H., D{\'e}barre, D., El Amri, N., Verdier, C., Richter, R. P., \& Bureau, L. (2018). Elastohydrodynamic lift at a soft wall. Physical Review Letters, 120(19), 198001.
		
		\bibitem{ding2025long}Ding, L. (2025). Long-time asymptotics of passive scalar transport in periodically modulated channels. Journal of Fluid Mechanics, 1023, A19. doi:10.1017/jfm.2025.10814.
		
		\bibitem{ding2023shear}Ding, L. (2023). Shear dispersion of multispecies electrolyte solutions in the channel domain. Journal of Fluid Mechanics, 970, A27
		
		\bibitem{wang2022reduced} Wang, X., \& Christov, I. C. (2022). Reduced modelling and global instability of finite-Reynolds-number flow in compliant rectangular channels. Journal of Fluid Mechanics, 950, A26.
		
		\bibitem{martinez2020start} Mart{\'\i}nez-Calvo, A., Sevilla, A., Peng, G. G., \& Stone, H. A. (2020). Start-up flow in shallow deformable microchannels. Journal of Fluid Mechanics, 885, A25.
		
		\bibitem{inamdar2020unsteady} Inamdar, T. C., Wang, X., \& Christov, I. C. (2020). Unsteady fluid-structure interactions in a soft-walled microchannel: A one-dimensional lubrication model for finite Reynolds number. Physical Review Fluids, 5(6), 064101.
		
		\bibitem{ramos2021fluid} Ramos-Arzola, L., \& Bautista, O. (2021). Fluid structure-interaction in a deformable microchannel conveying a viscoelastic fluid. Journal of Non-Newtonian Fluid Mechanics, 296, 104634.
		
		\bibitem{boyko2022flow} Boyko, E., Stone, H. A., \& Christov, I. C. (2022). Flow rate–pressure drop relation for deformable channels via fluidic and elastic reciprocal theorems. Physical Review Fluids, 7(9), L092201.
		
		\bibitem{gervais2006flow} Gervais, T., El-Ali, J., G{\"u}nther, A., \& Jensen, K. F. (2006). Flow-induced deformation of shallow microfluidic channels. Lab on a Chip, 6(4), 500-507.
		
		\bibitem{chun2025experimental} Chun, S., Christov, I. C., \& Feng, J. (2025). Experimental investigation of the flow rate–pressure drop relation of a viscoelastic Boger fluid in a deformable channel. Physical Review Applied, 24(3), 034001.
		
		\bibitem{keiser20222intermittent}Keiser, L., Marmottant, P., \& Dollet, B. (2022). Intermittent air invasion in pervaporating compliant microchannels. Journal of Fluid Mechanics, 948, A52.
		
		\bibitem{boyko2023non} Boyko, E., \& Christov, I. C. (2023). Non-Newtonian fluid–structure interaction: Flow of a viscoelastic Oldroyd-B fluid in a deformable channel. Journal of Non-Newtonian Fluid Mechanics, 313, 104990.
		
		\bibitem{chun2024flow}Chun, S., Boyko, E., Christov, I. C., \& Feng, J. (2024). Flow rate–pressure drop relations for shear-thinning fluids in deformable configurations: Theory and experiments. Physical Review Fluids, 9(4), 043302.
		
		\bibitem{boyko2025interplay} Boyko, E. (2025). Interplay between complex fluid rheology and wall compliance in the flow resistance of deformable axisymmetric configurations. Journal of Non-Newtonian Fluid Mechanics, 336, 105380.
		
		\bibitem{green2022effects}Green, Y. (2022). Effects of surface-charge regulation, convection, and slip lengths on the electrical conductance of charged nanopores. Physical Review Fluids, 7(1), 013702.
		
		\bibitem{chakraborty2010influence} Chakraborty, J., \& Chakraborty, S. (2010). Influence of streaming potential on the elastic response of a compliant microfluidic substrate subjected to dynamic loading. Physics of Fluids, 22(12).
		
		\bibitem{matse2018counterion} Matse, M., Berg, P., \& Eikerling, M. (2018). Counterion flow through a deformable and charged nanochannel. Physical Review E, 98(5), 053101.
		
		\bibitem{mandal2015electro}Mandal, S., Ghosh, U., Bandopadhyay, A., \& Chakraborty, S. (2015). Electro-osmosis of superimposed fluids in the presence of modulated charged surfaces in narrow confinements. Journal of Fluid Mechanics, 776, 390-429.
		
		\bibitem{pal2024electroosmotic}Pal, S. K., Mahapatra, P., Ohshima, H., \& Gopmandal, P. P. (2024). Electroosmotic flow modulation and enhanced mixing through a soft nanochannel with patterned wall charge and hydrodynamic slippage. Industrial \& Engineering Chemistry Research, 63(29), 12977-12998.
		
		\bibitem{mukherjee2022electrokinetically}Mukherjee, S., Dhar, J., DasGupta, S., \& Chakraborty, S. (2022). Electrokinetically augmented load bearing capacity of a deformable microfluidic channel. Physics of Fluids, 34(8).
		
		\bibitem{boyko2020interfacial}Boyko, E., Ilssar, D., Bercovici, M., \& Gat, A. D. (2020). Interfacial instability of thin films in soft microfluidic configurations actuated by electro-osmotic flow. Physical Review Fluids, 5(10), 104201.
		
		\bibitem{bhaskaran2024elasto}Bhaskaran, A. M., Agrawal, S., Sarkar, K., \& Dhar, P. (2024). Elasto-compliance of harmonically stimulated soft micro-gaps during electro-magneto-kinetic flows. Soft Matter, 20(30), 5969-5982.
		
		\bibitem{maroundik2025diffusioosmotic}Maroundik, N., Ilssar, D., \& Boyko, E. (2025). Diffusioosmotic flow in a soft microfluidic configuration induces fluid-structure instability. Physical Review Fluids, 10(10), 104203.
		
		\bibitem{mcnamee2019Effect} McNamee, C. E. (2019). Effect of a liquid flow on the forces between charged solid surfaces and the non-equilibrium electric double layer. Advances in Colloid and Interface Science, 266, 21-33.
		
		\bibitem{norouzisadeh2024a}Norouzisadeh, M., Leroy, P., \& Soulaine, C. (2024). A lubrication model with slope-dependent disjoining pressure for modeling wettability alteration. Computer Physics Communications, 298, 109114.
		
		\bibitem{yaros2003evaluation}Yaros, H. D., Newman, J., \& Radke, C. J. (2003). Evaluation of DLVO theory with disjoining-pressure and film-conductance measurements of common-black films stabilized with sodium dodecyl sulfate. Journal of colloid and interface science, 262(2), 442-455.
		
		\bibitem{rodriguez2025gaseous} Rodr{\'\i}guez, A., Arcos, J., Méndez, F., \& Bautista, O. (2025). Gaseous slip flow through a shallow deformable microchannel. Physics of Fluids, 37(6).

        \bibitem{israelachvili2011intermolecular}Israelachvili, J. N. (2011). Intermolecular and surface forces. Academic Press.

        \bibitem{leal2007advanced} Leal, L. G. (2007). Advanced transport phenomena: fluid mechanics and convective transport processes (Vol. 7). Cambridge University Press.
		
        \bibitem{timoshenko1959theory}Timoshenko, S., \& Woinowsky-Krieger, S. (1959). Theory of plates and shells.		
		
	\end{thebibliography}
\end{document}